\documentclass[conference]{IEEEtran}
\IEEEoverridecommandlockouts
\usepackage{amsmath,amssymb,amsfonts}
\usepackage{textcomp}
\usepackage{xcolor}
\def\BibTeX{{\rm B\kern-.05em{\sc i\kern-.025em b}\kern-.08em
    T\kern-.1667em\lower.7ex\hbox{E}\kern-.125emX}}

\usepackage{tikz}
\usepackage{booktabs} % For formal tables
\usepackage{filecontents}

\usepackage{mathptmx}

\usepackage{fancyhdr}
\usepackage[normalem]{ulem}
\usepackage{epsfig}
\usepackage{graphicx}
\usepackage{algorithm}
\usepackage{algorithmic}
\usepackage{amsmath}
\usepackage{setspace}
\usepackage{hyperref}
\usepackage{multirow}
\usepackage{booktabs}
\usepackage{flushend}
\renewcommand{\em}{\it}

\newcommand{\ignore}[1]{}

\def\cfigure[#1,#2,#3]{
\begin{figure}[t]
\begin{center}

\includegraphics[width=3.4in]{#1} 
 
\vspace*{-1mm}
\caption[]{#2
\vspace*{-0.1in}
} \label{#3}
 
\end{center}
\vspace*{-0.15in}
\end{figure}}

\def\cfiguretemp[#1,#2,#3]{
\begin{figure}
\vspace*{0mm}
\begin{center}

\includegraphics[width=3.5in]{#1} 
 
\vspace*{-3mm}
\caption[]{#2
} \label{#3}
 
\vspace*{-5mm}
\end{center}
\vspace*{-2mm}
\end{figure}}

\def\wfigure[#1,#2,#3]{
\begin{figure*}
\vspace*{0mm}
\begin{center}
 
\includegraphics[width=7in]{#1} 
 
\vspace*{-2mm}
\caption[]{#2
} 
\label{#3}
\end{center}
\vspace*{-0.2in}
\end{figure*}}

\def\threefigure[#1,#2,#3,#4,#5]{
\begin{figure*}
\vspace*{0mm}
\begin{center}

\begin{tabular}{ccc}
\includegraphics[width=2in]{#1} & \includegraphics[width=2in]{#2} &  \includegraphics[width=2in]{#3} \\
(a) & (b) & (c) \\
\end{tabular}

\vspace*{-3mm}\caption[]{#4
} \label{#5}

\vspace*{-5mm}
\end{center}
\vspace*{-2mm}
\end{figure*}}

\def\dcfigure[#1,#2,#3,#4,#5,#6]{
{
\begin{figure*}
\vspace*{0.2in}\
\begin{center}
\begin{minipage}[c]{3in}{
\includegraphics[width=3in]{#1} 
\vspace*{-3mm}\caption[]{#2} \label{#3} \
}\end{minipage}\hspace*{0.5in}\
\begin{minipage}[c]{3in}{
\includegraphics[width=3in]{#4} 
\vspace*{-3mm}\caption[]{#5}\label{#6} \
}\end{minipage}
\end{center}
\vspace*{-0.4in}\
\end{figure*}
}
}

\def\qcfigure[#1,#2,#3,#4,#5,#6]{
{
\begin{figure*}
\vspace*{0.2in}\
\begin{center}
\begin{minipage}[c]{3in}{
\includegraphics[width=3in]{#1} 
\vspace*{-3mm}
}
\end{minipage}\hspace*{0.5in}\
\begin{minipage}[c]{3in}{
\includegraphics[width=3in]{#2} 
\vspace*{-3mm}
}\end{minipage}

\begin{minipage}[c]{3in}{
\includegraphics[width=3in]{#3} 
\vspace*{-3mm}
}
\end{minipage}\hspace*{0.5in}\
\begin{minipage}[c]{3in}{
\includegraphics[width=3in]{#4} 
\vspace*{-3mm}
}\end{minipage}
\end{center}
\caption[]{#5}\label{#6}
\end{figure*}
}
}

\def\twfigureabc[#1,#2,#3,#4,#5]{
{
\tiny
\begin{figure}
\begin{center}
\begin{minipage}[c]{3.4in}{
\includegraphics[width=3.4in]{#1} 
\vspace*{-4mm}
}
\end{minipage}
\\(a)\\

\begin{minipage}[c]{3.4in}{
\includegraphics[width=3.4in]{#2} 
\vspace*{-4mm}
}\end{minipage}
\\(b)\\

\begin{minipage}[c]{3.4in}{
\includegraphics[width=3.4in]{#3} 
\vspace*{-4mm}
}
\end{minipage}
\\(c)\\
\end{center}
\vspace*{-4mm}
\caption[]{#4}
\label{#5}
\end{figure}
}
}

\def\twfigure[#1,#2,#3,#4,#5]{
{
\begin{figure}
\vspace*{0.2in}\
\begin{center}
\begin{minipage}[c]{6.5in}{
\includegraphics[width=6.5in]{#1} 
\vspace*{-3mm}
}
\end{minipage}

\begin{minipage}[c]{6.5in}{
\includegraphics[width=6.5in]{#2} 
\vspace*{-3mm}
}\end{minipage}

\begin{minipage}[c]{6.5in}{
\includegraphics[width=6.5in]{#3} 
\vspace*{-3mm}
}
\end{minipage}
\end{center}
\caption[]{#4}\label{#5}
\end{figure}
}
}

\def\dwfigure[#1,#2,#3,#4]{
{
\begin{figure*}
\vspace*{0.2in}\
\begin{center}
\begin{minipage}[c]{3.5in}{
\includegraphics[width=3.5in]{#1} 
\vspace*{-3mm}
}
\end{minipage}

\begin{minipage}[c]{3.5in}{
\includegraphics[width=3.5in]{#2} 
\vspace*{-3mm}
}\end{minipage}

\end{center}
\caption[]{#3}\label{#4}
\end{figure*}
}
}

\def\dssfigure[#1,#2,#3,#4,#5,#6]{
{
\begin{figure*}
\vspace*{0.2in}\
\begin{center}
\begin{minipage}[c]{4in}{
\includegraphics[width=4in]{#1}
\vspace*{-3mm}\caption[]{#2} \label{#3} \
}\end{minipage}\hspace*{0.5in}\
\begin{minipage}[c]{2in}{
\includegraphics[width=2in]{#4}
\vspace*{-3mm}\caption[]{#5}\label{#6} \
}\end{minipage}
\end{center}
\vspace*{-0.4in}\
\end{figure*}
}
}

\def\dsfigure[#1,#2,#3,#4,#5,#6]{
{
\begin{figure*}
\vspace*{0.2in}\
\begin{center}
\begin{minipage}[c]{3in}{
\includegraphics[width=3in]{#1}
\vspace*{-3mm}\caption[]{#2} \label{#3} \
}\end{minipage}\hspace*{0.5in}\
\begin{minipage}[c]{3in}{
\hspace*{0.5in}\
\includegraphics[height=3in]{#4}
\vspace*{-3mm}\caption[]{#5}\label{#6} \
}\end{minipage}
\end{center}
\vspace*{-0.4in}\
\end{figure*}
}
}

\def\dsyfigure[#1,#2,#3,#4,#5,#6]{
{
\begin{figure*}
\vspace*{0.2in}\
\begin{center}
\begin{minipage}[c]{2.5in}{
\includegraphics[height=2.5in]{#1}
\vspace*{-3mm}\caption[]{#2} \label{#3} \
}\end{minipage}\hspace*{0.5in}\
\begin{minipage}[c]{2.5in}{
\includegraphics[height=2.5in]{#4}
\vspace*{-3mm}\caption[]{#5}\label{#6} \
}\end{minipage}
\end{center}
\vspace*{-0.4in}\
\end{figure*}
}
}

\def\dyfigure[#1,#2,#3,#4,#5,#6]{
{
\begin{figure*}
\vspace*{0.2in}\
\begin{center}
\begin{minipage}[c]{3in}{
\includegraphics[height=3in]{#1} 
\vspace*{-3mm}\caption[]{#2} \label{#3} \
}\end{minipage}\hspace*{0.5in}\
\begin{minipage}[c]{3in}{
\includegraphics[height=3in]{#4} 
\vspace*{-3mm}\caption[]{#5}\label{#6} \
}\end{minipage}
\end{center}
\vspace*{-0.4in}\
\end{figure*}
}
}

\def\dyoldfigure[#1,#2,#3,#4,#5,#6]{
{
\begin{figure*}
\vspace*{0.2in}\
\begin{center}
\begin{minipage}[c]{3in}{
\epsfysize=2.0in\
\hspace{0.5in}\
\epsfbox{#1}
\vspace*{-3mm}\caption[]{#2} \label{#3} \
}\end{minipage}\hspace*{0.25in}\
\begin{minipage}[c]{3in}{
\epsfysize=2.0in\
\hspace{0.5in}\
\epsfbox{#4}
\vspace*{-3mm}\caption[]{#5}\label{#6} \
}\end{minipage}
\end{center}
\vspace*{-0.4in}\
\end{figure*}
}
}

\def\cfiguredouble[#1,#2,#3,#4]{
\begin{figure}[t]
\begin{center}
\begin{minipage}[c]{3.4in}{
\epsfxsize=3.4in\
\epsfbox{#1}
}\end{minipage}\\(a)\\
\begin{minipage}[c]{3.4in}{
\epsfxsize=3.4in\
\epsfbox{#2}
}\end{minipage}\\(b)
\vspace*{-0.1in} 
\caption[]{#3}\label{#4}
\end{center}
\vspace*{-0.2in}\
\end{figure}
}

\def\wpfigure[#1,#2,#3,#4]{
\begin{figure*}
\vspace*{4mm}
\begin{center}

\includegraphics[width=#4]{#1} 

\vspace*{-3mm}\caption[]{#2
} \label{#3}

\vspace*{-5mm}
\end{center}
\end{figure*}}

\def\wprfigure[#1,#2,#3,#4,#5]{
\begin{figure*}
\vspace*{4mm}
\begin{center}

\includegraphics[width=#4, angle=#5]{#1} 

\vspace*{-3mm}\caption[]{#2
} \label{#3}

\vspace*{-5mm}
\end{center}
\end{figure*}}

\def\DoubleFigureWSlide[#1,#2,#3,#4,#5,#6,#7,#8,#9]{
\begin{figure*}
\vspace*{#9}
\begin{center}
\begin{minipage}{#4}
\includegraphics[width=#4]{#1}
\vspace*{-3mm}\caption{#2
}\label{#3}
\end{minipage}
\hspace{2em}
\begin{minipage}{#8}
\includegraphics[width=#8]{#5}
\vspace*{-3mm}\caption{#6
}\label{#7}
\end{minipage}
\vspace*{-5mm}
\end{center}
\end{figure*}
}

\def\DoubleFigureW[#1,#2,#3,#4,#5,#6,#7,#8]{
\begin{figure*}
\vspace*{0in}
\begin{center}
\begin{minipage}{#4}
\includegraphics[width=#4]{#1}
\vspace*{-3mm}\caption{#2
}\label{#3}
\end{minipage}
\hspace{2em}
\begin{minipage}{#8}
\includegraphics[width=#8]{#5}
\vspace*{-3mm}\caption{#6
}\label{#7}
\end{minipage}
\vspace*{-5mm}
\end{center}
\end{figure*}
}

\def\DoubleFigureWHack[#1,#2,#3,#4,#5,#6,#7,#8]{
\begin{figure*}
\vspace*{0in}
\begin{center}
\begin{minipage}{3in}
\includegraphics[width=#4]{#1}
\vspace*{-3mm}\caption{#2
}\label{#3}
\end{minipage}
\hspace{2em}
\begin{minipage}{3in}
\includegraphics[width=#8]{#5}
\vspace*{-3mm}\caption{#6
}\label{#7}
\end{minipage}
\vspace*{-5mm}
\end{center}
\end{figure*}
}

\def\ddcfigure[#1,#2,#3,#4]{
\begin{figure*}[t]
\vspace*{0.2in}\
\begin{center}
\begin{minipage}[c]{3in}{
\includegraphics[width=3in]{#1} 
}\end{minipage}\hspace{0.5in}\
\begin{minipage}[c]{3in}{
\includegraphics[width=3in]{#2} 
}\end{minipage}\vspace*{-0.10in} \caption[]{#3}\label{#4}
\end{center}
\vspace*{-0.4in}\
\end{figure*}
}

\def\dddcfigure[#1,#2,#3,#4]{
\begin{figure*}
\vspace*{0.2in}\
\begin{center}
\begin{tabular}{cc}
\includegraphics[width=3in]{#1} &
\includegraphics[width=3in]{#2} \\
(a) & (b) \\
\end{tabular}\vspace*{-0.10in}\caption[]{#3}\label{#4}
\end{center}
\vspace*{-0.4in}\
\end{figure*}
}

\def\qqcfigure[#1,#2,#3,#4,#5,#6]{
\begin{figure*}[t]
\begin{center}
\begin{tabular}{cc}
\includegraphics[width=3.3in]{#1} &
\includegraphics[width=3.3in]{#2} \\
(a) & (b) \\
\includegraphics[width=3.3in]{#3} &
\includegraphics[width=3.3in]{#4} \\
(c) & (d) \\
\end{tabular}\vspace*{-0.10in}\caption[]{#5}\label{#6}
\end{center}
\vspace*{-0.4in}\
\end{figure*}
}
\def\qqcfiguresinglecol[#1,#2,#3,#4,#5,#6]{
\begin{figure}[t]
\begin{center}
\begin{tabular}{cc}
\hspace*{-0.2in}\includegraphics[width=2in]{#1} &
\hspace*{-0.4in}\includegraphics[width=2in]{#2} \\
\hspace*{-0.2in}(a) & 
\hspace*{-0.4in}(b) \\
\hspace*{-0.2in}\includegraphics[width=2in]{#3} &
\hspace*{-0.2in}\includegraphics[width=2in]{#4} \\
\hspace*{-0.2in}(c) & 
\hspace*{-0.4in}(d) \\
\end{tabular}\vspace*{-0.10in}\caption[]{#5}\label{#6}
\end{center}
\vspace*{-0.4in}\
\end{figure}
}

\def\qqcfigureInAColumn[#1,#2,#3,#4,#5,#6]{
\begin{figure}[t]
\begin{center}
\includegraphics[width=3.3in]{#1} \\
(a)\\
\includegraphics[width=3.3in]{#2} \\
(b) \\
\includegraphics[width=3.3in]{#3} \\
(c) \\
\includegraphics[width=3.3in]{#4} \\
(d) \\
\vspace*{-0.10in}\caption[]{#5}\label{#6}
\end{center}
\vspace*{-0.5in}\
\end{figure}
}

\def\sixfigure[#1,#2,#3,#4,#5,#6,#7,#8]{
\begin{figure*}[t]
\vspace*{0.2in}\
\begin{center}
\begin{tabular}{cc}
\includegraphics[width=3in]{#1} &
\includegraphics[width=3in]{#2} \\
(a) & (b) \\
\includegraphics[width=3in]{#3} &
\includegraphics[width=3in]{#4} \\
(c) & (d) \\
\includegraphics[width=3in]{#5} &
\includegraphics[width=3in]{#6} \\
(e) & (f) \\
\end{tabular}\vspace*{-0.10in}\caption[]{#7}\label{#8}
\end{center}
\vspace*{-0.4in}\
\end{figure*}
}

\def\ddcfigureSlide[#1,#2,#3,#4,#5]{
\begin{figure*}
\vspace*{#5}\
\begin{center}
\begin{minipage}[c]{3in}{
\includegraphics[height=3in]{#1} 
}\end{minipage}\hspace{0.5in}\
\begin{minipage}[c]{3in}{
\includegraphics[height=3in]{#2} 
}\end{minipage}\vspace*{-0.10in} \caption[]{#3}\label{#4}
\end{center}
\vspace*{-0.4in}\
\end{figure*}
}

\def\dcfigureSingleCol[#1,#2,#3,#4]{
\begin{figure}
\begin{center}
\begin{tabular}{cc}
\hspace*{-0.2in}\includegraphics[width=2.4in]{#2} &
\hspace*{-0.2in}\includegraphics[width=1.2in]{#1} \\
\hspace*{-0.2in}(a) & 
\hspace*{-0.2in}(b) \\
\end{tabular}\vspace*{-0.10in}\caption[]{#3}\label{#4}
\end{center}
\vspace*{-0.4in}\
\end{figure}
}

\def\cxfigure[#1,#2,#3]{
\begin{figure}
\vspace*{4mm}
\begin{center}
 
\epsfxsize=2.5in\
\epsfbox{#1}\
 
\vspace*{-0.10in}\caption[]{#2
} \label{#3}
 
\vspace*{-5mm}
\end{center}
\vspace*{-2mm}
\end{figure}}

\newcommand{\x}{$\times$}

\newif\ifremark
\long\def\remark#1{
\ifremark%
        \begingroup%
        \dimen0=\columnwidth
        \advance\dimen0 by -1in%
        \setbox0=\hbox{\parbox[b]{\dimen0}{\protect\em #1}}
        \dimen1=\ht0\advance\dimen1 by 2pt%
        \dimen2=\dp0\advance\dimen2 by 2pt%
        \vskip 0.25pt%
        \hbox to \columnwidth{%
                \vrule height\dimen1 width 3pt depth\dimen2%
                \hss\copy0\hss%
                \vrule height\dimen1 width 3pt depth\dimen2%
        }%
        \endgroup%
\fi}

\remarktrue

\begin{document}
%\pagestyle{plain}

%\input{summary}
%\clearpage
%\pagestyle{myheadings}
%\pagenumbering{arabic}

%\newenvironment{smallitemize}{\begin{list}{$\bullet$}{\topsep0.0in\itemsep0.0in\parsep0.0in\partopsep0.0in\itemindent0.1in\leftmargin0.1in}}{\end{list}}
%\newenvironment{smallitemize}{\begin{list}{$\bullet$}{\topsep0.1in\itemsep0.1in\parsep0.1in\partopsep0.1in\itemindent0.1in\leftmargin0.1in}}{\end{list}}
%\newenvironment{smallitemize}{\begin{list}{$\bullet$}{\topsep0.05in\itemsep0.05in\parsep0.0in\partopsep0.05in\itemindent0.05in\leftmargin0.05in}}{\end{list}}
%{\topsep{0in}\itemsep{0.1in}\itemindent{0.1in}}
%%%%%%%%%%%%%%%%%%%%%%%%

\newcommand{\mm}{mm$^2$}
\newcommand{\figtitle}[1]{\textbf{#1}}
\newcommand{\us}{$\mu$s}
\newcommand{\fixme}[1]{#1}
\newcommand{\adrian}[1]{{\color{green}\textbf{#1}}}
\newcommand{\laura}[1]{{\color{pink}\textbf{#1}}}
\newcommand{\joel}[1]{{\color{red}\textbf{#1}}}
\newcommand{\ameen}[1]{{\color{blue}\textbf{#1}}}
\newcommand{\arup}[1]{{\color{yellow}\textbf{#1}}}
\newcommand{\hungwei}[1]{{\color{blue}{#1}}}
\newcommand{\otto}[1]{{\color{red}\textbf{#1}}}
\newcommand{\tei}[1]{{\color{green}\textbf{#1}}}
\newcommand{\Bella}[1]{{\color{blue}\textbf{\textit{#1}}}}
\newcommand{\kathy}[1]{{\color{cyan}\textbf{\textit{#1}}}}
%BEGIN: CMF DEFINITIONS:
%\newcommand{\CMF}[1]{{\color{orange}{#1}}} %For substitutions
\newcommand{\CMF}[1]{{{#1}}} %For substitutions
\newcommand{\CMFdel}[1]{} %For deletions
\newcommand{\CMFfin}[1]{#1}
\newcommand{\CMFcom}[1]{}
%END: CMF DEFINITIONS
%BEGIN: CMF DEFINITIONS FOR SECOND REVISIONS (SR; MAR-2020):
\newcommand{\CMFSR}[1]{{\color{orange}{#1}}} %For insertions
\newcommand{\CMFSRdel}[1]{} %For deletions
\newcommand{\CMFSRcom}[1]{{\color{orange}\textbf{\textit{[CMF: #1]}}}} %For comments
%END: CMF DEFINITIONS FOR SECOND REVISIONS (MAR-2020):

\newcommand{\hungweinote}[1]{{\color{blue}\textbf{#1}}}

\newcommand{\ottodone}[1]{}
\newcommand{\kathydone}[1]{}

\newcommand{\note}[2]{{\color{red}\fixme{$\ll$ #1 $\gg$ #2}}}
\newcommand{\myitem}[1]{\hspace*{-\parindent}\textbf{#1}\hspace*{\parindent}}
\newcommand{\OpenWVR}[1]{OpenUVR}
\newcommand{\OpenUVR}[1]{OpenUVR}
\newcommand{\BaseUVR}[1]{HWA-Gaming\-Anywhere}
\newcommand{\OPENUVR}[1]{OPENUVR}
%BEGIN: CMF REVISIONS TO tikz COMMANDS
%original command:
%\newcommand*\circled[1]{\tikz[baseline=(char.base)]{
%            \node[shape=circle,fill,inner sep=1pt] (char)
%{\textcolor{white}{#1}};}}

%modified command (minimum size=4.85mm,font=\small):
\newcommand*\circled[1]{\tikz[baseline=(char.base)]{
            \node[shape=circle,fill,inner sep=1pt,minimum size=4.85mm,font=\small] (char)
{\textcolor{white}{#1}};}}

%original command:
%\newcommand*\circlel[1]{\tikz[baseline=(char.base)]{
%            \node[shape=circle,draw,inner sep=1pt] (char)
%{\textcolor{black}{#1}};}}

%modified command (minimum size=4.85mm,font=\small):
\newcommand*\circlel[1]{\tikz[baseline=(char.base)]{
            \node[shape=circle,draw,inner sep=1pt,minimum size=4.85mm,font=\small] (char)
{\textcolor{black}{#1}};}}

%END: CMF REVISIONS TO tikz COMMANDS
%\let\hacksection\section
%\renewcommand{\section}{\vspace{-1ex}\hacksection}
%\let\hacksubsection\subsection
%\renewcommand{\subsection}{\vspace{-1ex}\hacksubsection}
%\let\hacksubsubsection\subsubsection
%\renewcommand{\subsubsection}{\vspace{-1ex}\hacksubsubsection}

%\title{\OpenWVR{}: an Open-Source Framework for Untethered Virtual Reality Systems}
\title{\OpenWVR{}: an Open-Source System Framework for Untethered Virtual Reality Applications
}
%\title{\OpenWVR{}: reducing latency and preserving quality in untethered real-time virtual-reality gaming platforms}

%Eliminating Redundant Memory Copies for Untethered Virtual Reality Applications}
%\author{Submitted for blind review}
%\vspace{-12ex}
\author{
\IEEEauthorblockN{Alec Rohloff\textsuperscript{\textsection}}
\IEEEauthorblockA{Applied Research Associates}
\and
\IEEEauthorblockN{Zackary Allen\textsuperscript{\textsection}}
\IEEEauthorblockA{IBM}
\and
\IEEEauthorblockN{Kung-Min Lin\textsuperscript{\textsection}}
\IEEEauthorblockA{University of California, Berkeley}
\and
\IEEEauthorblockN{Joshua Okrend\textsuperscript{\textsection}}
\IEEEauthorblockA{Riverside Technology, Inc.}
\and
\IEEEauthorblockN{Chengyi Nie\textsuperscript{\textsection}}
\IEEEauthorblockA{Stony Brook University, New York}
\and
\IEEEauthorblockN{Yu-Chia Liu}
\IEEEauthorblockA{University of California, Riverside}
\and
\IEEEauthorblockN{Hung-Wei Tseng}
\IEEEauthorblockA{University of California, Riverside}
}
\date{}

\maketitle
\footnote{This paper is a pre-print of a paper in the 27th IEEE Real-Time and Embedded 
Technology and Applications Symposium. Please refer to the conference proceedings
for the most complete version.}
\begingroup\renewcommand\thefootnote{\textsection}
\footnotetext{Alec Rohloff, Zackary Allen and Joshua Okrend are former
undergraduate researchers at North Carolina State University advised by
Hung-Wei Tseng. Kung-Min Lin
is a former summer intern at University of California, Riverside  advised by
Hung-Wei Tseng. Chengyi Nie
is a former undergraduate researcher at University of California, Riverside  advised by
Hung-Wei Tseng.}
\begin{abstract}
Advancements in heterogeneous computing
technologies enable the significant potential of virtual
reality (VR) applications. To offer the best user experience (UX),
a system
should adopt an untethered, wireless-network-based architecture to
transfer VR content between the user and the content generator.
However, modern wireless network technologies
make implementing such an
architecture challenging, as VR applications require superior video quality---with high
resolution, high frame rates, and very low latency.

This paper presents \OpenWVR{}, an open-source framework that
uses commodity hardware components to satisfy
the demands of interactive, real-time VR applications. \OpenWVR{} 
significantly improves UX through a redesign
of the system stack and addresses the most
time-sensitive issues associated with
redundant memory copying in modern
computing systems. \OpenWVR{} presents a cross-layered VR datapath to avoid 
redundant data operations and computation among system components,
\OpenWVR{} customizes the network stack to eliminate unnecessary
memory operations incurred by mismatching data formats in each layer, and 
\OpenWVR{} uses feedback from mobile devices to remove memory buffers.

Together, these modifications allow \OpenWVR{}
to reduce VR application delays to 14.32~ms, meeting the 20~ms minimum latency
in avoiding motion sickness. As an
open-source system that is fully compatible with
commodity hardware,
\OpenWVR{} offers the research
community an opportunity to develop, investigate,
and optimize applications for untethered, high-performance
VR architectures.
\end{abstract}

\section{Introduction}
\label{sec:introduction}
With high-performance graphics processing units (GPUs) and
hardware accelerators, modern heterogeneous personal computers already have the computing power
needed for virtual reality (VR) applications.
But latency and mobility are still a problem; to provide
the best user experience (UX), a system must be enabling the user to move around freely
and able to deliver VR content
to the user in 20~ms to avoid motion sickness in real-time gaming~\cite{ShootingAMovingTarget, Furion, FlashBackBoos, MotionSickness,
doi:10.1177/0018720811428734, 8329628, steam, Cybersickness,
MeaureVRLatency, 10.1145/3131214.3131218}.
%with minimal delay while enabling the user to move around freely.
Unfortunately, existing commercialized VR solutions adopt either a 
tethered architecture that limits user mobility~\cite{Oculus, VIVE} or an
untethered architecture (using mobile devices) that limits visual effects~\cite{Daydream,GearVR,OculusGo,LenovoMirageSolo}.

%Conventional wisdom says that 
The major obstacle in developing effective untethered VR (UVR) systems lies with a mismatch
between the bandwidth demand for high-quality visual
content and the sustainable bandwidth
provided by the underlying wireless VR links. 
Though compression helps lower the bandwidth demand to tens of Mbps,
the increased visual latency is unacceptable to latency-sensitive VR
applications.
\ignore{
Such latency is less problematic for VR applications
with slow, predictable user movements but
leads to motion sickness in real-time gaming if the latency is more than
%CMFSR: CHANGED "hootingAMovingTarget" TO "ShootingAMovingTarget" IN THE CITATIONS BELOW
20~ms~\cite{ShootingAMovingTarget, Furion, FlashBackBoos, MotionSickness,
doi:10.1177/0018720811428734, 8329628, steam, Cybersickness,
MeaureVRLatency, 10.1145/3131214.3131218}.}
%~\cite{FlashBackBoos, MotionSickness}.
Recent research projects
have promoted millimeter-wave (mmWave) wireless
technologies to avoid the trade-off between latency and
bandwidth~\cite{TPCast, CuttingtheCord, MoVR}. However, such technologies severely
limit user mobility because they depend on line-of-sight signaling. As a result,
currently available UVR products can offer only mildly entertaining
novelty titles.
% instead of fully immersive experiences.

This paper presents \OpenWVR{}, a fully open-source, open-architecture
VR system framework that resolves the
three-way trade-off between latency, bandwidth, and UX.
Rather than treating
a VR application as a type of video streaming with bandwidth issues, as is often done,
\OpenWVR{} improves the end-to-end
latency that is critical to real-time, interactive action titles.
\OpenWVR{} also builds on an important lesson
from modern high-performance I/O systems---that redundant memory
copies comprise the most significant part of data-exchange latency~\cite{IX, MonetaD}. We have
found memory copies from three sources to be especially problematic for UVR architectures: (1) data exchanges
between memory spaces, (2) data transformations between system
modules/stacks, and (3) data buffering between system components.

\OpenWVR{}'s design revisits the complete datapath (including graphics/video libraries and the network stack), and accounts for
application demands in order to minimize the number
of memory copies in delivering VR content. The \OpenWVR{} API and kernel
modules work with commodity hardware
components to eliminate redundant memory buffers/operations and avoid context switches before
data is sent to or received from network stacks.
\OpenWVR{} further simplifies the network stack by using peer-to-peer
network architecture and abandoning the traditional multilayer
network-stack delivery design that regenerates packets to accommodate the headers of each layer.

Eliminating unnecessary memory copies
allows \OpenWVR{} to achieve an imperceptible delay of just 14.32~ms
 in real-time VR 
applications---and this is done with affordable components and
current-generation
Wi-Fi technologies. And because the \OpenUVR{} mobile user device (MUD) performs minimal computations, 
it can be powered by a simple Raspberry Pi single-board computer
that costs only USD 35.

By exploring the unique properties, challenges, and trade-offs inherent in
the design of a UVR system, this paper makes
several contributions.

\myitem{(1) The paper reveals the
inefficiencies of system stacks and the need for cross-layer optimizations in
wireless VR systems.} As an
experimental system, \OpenWVR{} shows that
the latency overhead from each layer in the system
stack can be remedied through optimal buffering, memory
copying, transcoding, and data exchange.
%The initial implementation of the system indicates that even with optimized
%hardware/software components,
Our observations suggest
that an efficient VR system must present local optimizations in all layers
instead of focusing on a single-point design.

\myitem{(2) The paper presents a
general design philosophy for wireless networking applications
that require low latency and high throughput.} \OpenWVR{}'s holistic, latency-mitigating 
system framework bypasses the inefficiencies of network stacks and
uses existing hardware accelerators
to yield a robust wireless VR infrastructure.
The same efficient design can easily be applied to
wireless systems with similar demands
as an alternative to expensive,
throughput-oriented wireless network technologies.

\myitem{(3) The paper advances research by
providing a publicly available, open-source solution to VR system challenges.}
\OpenWVR{}'s hardware components and
wireless technologies are widely available and relatively low-cost, and its software may be
freely inspected and modified. As a fully open-source solution, \OpenWVR{}
will help researchers extend and optimize VR architectures.~\footnote{We will make our
private GitHub repository public once the paper has been accepted.}
\ignore{
The rest of this paper is organized as follows:
Section~\ref{sec:background} introduces the background of modern VR systems.
Section~\ref{sec:arch} provides an overview of the \OpenWVR{} system architecture.
Section~\ref{sec:methodology} describes our experimental platform.
Section~\ref{sec:result} presents our design space exploration on the basic
\OpenWVR{}.
Section~\ref{sec:system} depicts the optimizations in \OpenWVR{}.
%Section~\ref{sec:programming_model} depicts the programming model.
Section~\ref{sec:user_study} describes our user study.
Section~\ref{sec:discussions} discusses other factors that potentially
affect the performance of \OpenWVR{} and compares \OpenWVR{} with similar
counterparts.
%improves the performance of \KaleidoSSD{}.
Section~\ref{sec:related_works} provides a summary of related work to put
this project in context, and
Section~\ref{sec:conclude} concludes the paper.
%Section numbers don't populate accurately in the PDF. Please review.
}

\section{Background}
\label{sec:background}
We now describe the basic architecture of a VR system and the challenges of
building a VR system that promotes UX. We also discuss existing VR solutions and
their associated drawbacks.

\subsection{Basic operations and challenges of VR systems}
\label{sec:vr}
%\cfigure[Figures/VR.pdf, {The \CMFdel{``loop'' of }VR task\CMFdel{s}\CMF{ loop}},fig:vr_loop]
A typical VR system needs to accomplish three main types of tasks:

\myitem{User input collection} The system must gather all user
inputs from the commands that users explicitly send through
a device (e.g., a keyboard, microphone, or
controller) and data from
sensors that detect the user's motions.

\myitem{Content generation} The system must process user inputs, select the
elements to appear in the VR
content, and render the content according to application
specifications. Rendering VR content requires high-performance computing resources
and data retrieval from either local storage or a cloud service.

\myitem{Content presentation} The system must display the VR
content to the user,
usually as video frames and sound
effects. The user then
reacts to the VR content and generates
corresponding inputs that initiate another
task loop.

To deliver the best UX,
a system needs to address three challenges when
completing the aforementioned VR tasks:

\myitem{Visual effects} The system must render video frames that
provide immersive views of virtual scenes. Such rendering
requires high resolutions and high
frame rates (on the order of 60 frames per second [FPS] in 1080p resolution
for both eyes)~\cite{VRChallenges}.

\myitem{Response latency} A delayed response in immersive
VR applications can cause motion
sickness, so a VR system must complete all task loops with reasonably low latency
(20~ms).

%~\cite{Steam}.
\myitem{Physical comfort} In many VR applications, especially games, the
user can move about to perform VR
actions, so the MUD, which sends/receives
inputs and displays visual effects, must be lightweight and untethered to maximize comfort and minimize constraints.

Unfortunately, no existing commercial VR
system rises to meet all of the above challenges.

\subsection{Existing VR systems}
\label{sec:background_VR_systems}
\twfigureabc[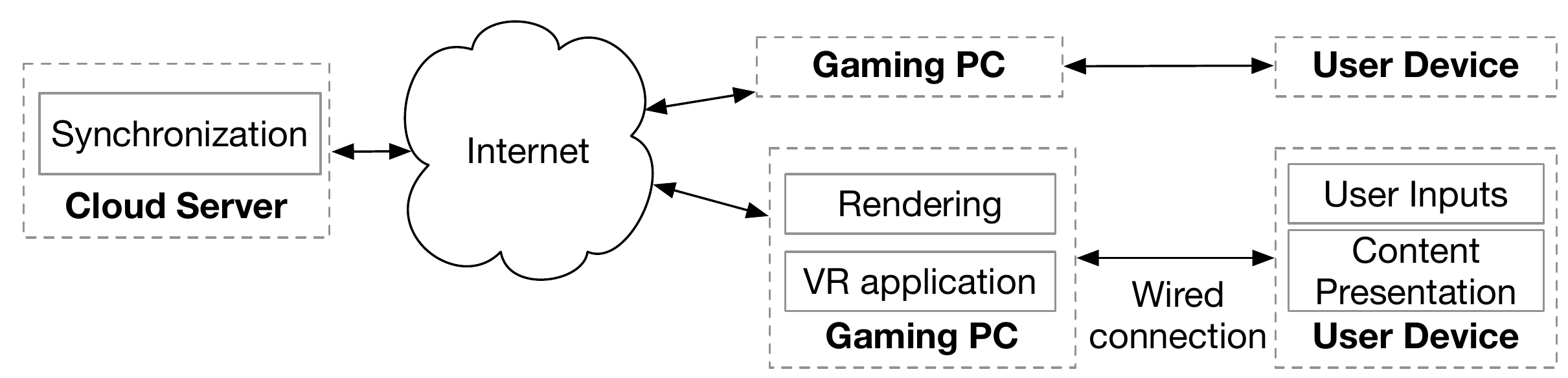,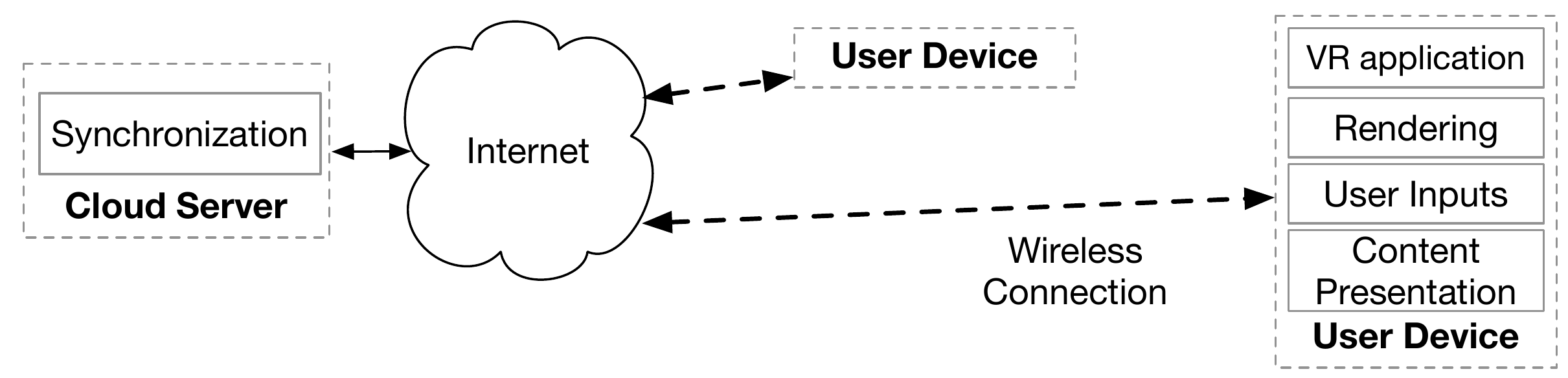,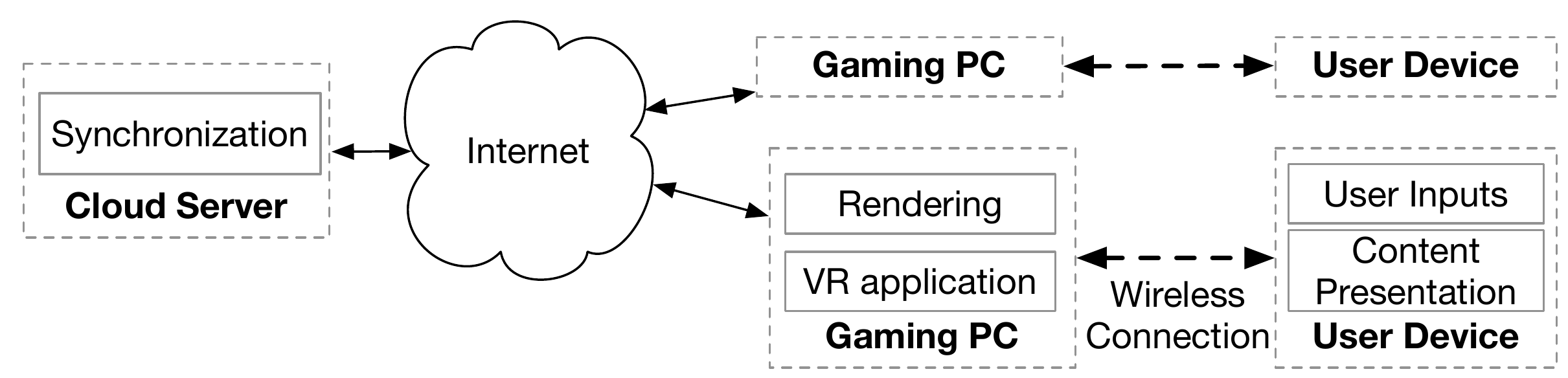, {The VR task allocations in
different VR system architectures},fig:vr_loop]

Since the release of Facebook's Oculus Rift VR headset in 2016~\cite{Oculus}, VR systems have steadily
gained market share as companies have introduced improved
designs. Figure~\ref{fig:vr_loop} depicts the three most common designs for modern VR systems
and how these systems allocate VR tasks among their components.

\myitem{Tethered device} Figure~\ref{fig:vr_loop}(a) illustrates the system architecture of a 
VR system that uses a tethered device. A VR system relying on
a tethered device consists of a high-performance PC and a low-performance wearable device.
These two components are connected by wires that power
the wearable device, transmit audio-visual output to the device, and
carry user input back to the PC from the device.
The Oculus Rift and the HTC Vive~\cite{VIVE} are
examples of such systems. The Rift only tracks the orientation of the
headset, and the Vive uses room sensors to track the motion of the user in 3D space.
In both systems, wires connected to the interactive device limit
the user's movements, create a tripping hazard, and undermine the immersive experience
at the heart of VR. 

\myitem{Untethered device without a host PC}
Figure~\ref{fig:vr_loop}(b) illustrates the concept behind this type of VR system.
All-in-one devices place all VR tasks on the MUD, with the exception of social-synchronization tasks
or leverage the cloud for rendering.
MUDs fall into two main categories:
MUDs powered by smartphones and MUDs powered by standalone devices. 

\ignore{
\CMF{(1)}\CMFdel{providing} headsets that allow smartphones
to act as VR devices, \CMF{(2)}\CMFdel{producing} standalone \CMF{VR }devices,
and \CMF{(3)} wirelessly stream\CMFdel{ing}\CMF{ed}, PC-rendered visuals.
}
Smartphone-powered headsets such as the Google Daydream~\cite{Daydream}
and the Samsung Gear VR~\cite{GearVR} rely on a smartphone's computing power 
to render images, the phone's screen to display the images, 
and the phone's accelerometers to track the user's motions. Since the
Google and Samsung systems are limited by the capabilities of the
user's smartphone, highly detailed VR games and experiences available on a 
PC are unavailable to users with low-end phones.
% Running an application on one of these VR systems
%also consumes energy on a mobile phone faster than in a general use case---the Google Pixel XL's battery 
%(3,450 mAh) is completely drained in about 3 hours \CMFSRcom{Citation?}. 

Standalone VR systems like the Oculus Go~\cite{OculusGo}, the Oculus Quest, and the Lenovo Mirage
Solo~\cite{LenovoMirageSolo} are similar
to smartphone-dependent VR devices, except that the Go and Solo have their own screens,
sensors, and smartphone-grade processors to render and display images.
As with the smartphone-dependent devices,
the Go and Solo are unable to render highly detailed, complex VR experiences
because the processors in their devices are so limited.

To enable complex VR experiences while maintaining the simplicity of MUDs,
the recent cloud services including Google Stadia~\cite{stadia}, Steam
Remote Play~\cite{SteamRemotePlay}, and NVIDIA's Geforce Now, adopts the
system architecture of using the GPUs on cloud servers for rendering. To
address the high latency due to the distance and unpredictable traffic
condiction between cloud servers and the MUD, the system intensively
employs techniques including post-render
warp~\cite{10.1145/253284.253292, 10.1145/3406187} that render only informative
parts of the upcoming frames on the remote cloud. However, doing this
requires significantly tuning on the game engine and the visual library,
increasing the overhead of game development while the MUD still suffers from
the long latency with cloud servers. 

\myitem{Untethered device with a host PC}
The most promising approach to creating a fully immersive wireless experience
on par with that of wired VR is to render visuals on a separate, powerful
computer and stream the visuals to a wearable device using a fast wireless network
(e.g., a network based on IEEE 802.11ac or 60~GHz mmWave solutions). 

A few startup companies like Amimon, KwikVR, and TPCAST (and some research groups as well) are developing 
such UVR systems using 60~GHz mmWave wireless networks; however, the systems haven't yet reached the wider
market due to the limitations of mmWave wireless technologies.
TPCAST advertises a latency of less than 10~ms using a
60~GHz wireless connection, but a 60 GHz signal suffers from very high
attenuation. Furthermore, a 60 GHz signal may rely on beam focusing to
overcome attenuation, so users can easily
lose the wireless connection for a VR application when they move the
head-mounted display (HMD) outside of the transmission beam. 
MoVR~\cite{MoVR} tries to address the limitation of 60~GHz mmWave with a specially designed antenna and 
mirrors carefully positioned in the VR environment; this is far from ideal, as MoVR still requires special
environment to work.  Zhong
et. al.~\cite{ZhongWiGigGaming} and Liu et. al.~\cite{CuttingtheCord} investigate the use of
60~GHz wireless networks and rendering pipelines placed between the
host PC and the receiver, but 60~GHz wireless networks require a high
decoding rate and have a severely limited range. These UVR systems also rely 
on laptop PCs that fail to match the power, weight, and budget limitations of a MUD. 

As with \OpenUVR{}, Furion~\cite{Furion} leverages mature commodity IEEE 802.11ac technology 
to transmit content from the host PC. By parallelizing the video codec and splitting 
background/foreground rendering, Furion minimizes latency to the
same level as \OpenUVR{}. However, Furion's approach requires multiple fully active
processor cores on both the host PC and MUD. As a result,
a UVR system with Furion still needs a smartphone 
%CMFSR: CHANGED "GamingAnywhere" TO "Gaming\-Anywhere" IN ORDER TO ENSURE PREFERRED HYPHENATION ACROSS LINE BREAKS
as a MUD. Gaming\-Anywhere~\cite{GamingAnywhere}
is an earlier open-source game-streaming system design that can be
used directly with an IEEE 802.11ac wireless network.
Like Furion, Gaming\-Anywhere requires a MUD with significant computing
power. Nonetheless, Gaming\-Anywhere is an open\CMFSR{-}source project, and the system's 
inefficiencies have benefitted \OpenUVR{}'s design.

%\input{model}
%\input{overview}
%CMF: ALL OCCURRENCES OF "GamingAnywhere" HAVE BEEN REPLACED WITH "Gaming\-Anywhere" TO ENSURE PREFERRED HYPHENATION FOR LINE BREAKS.
\section{Performance of Modern UVR Systems}
%\CMFSRcom{I changed ``untethered VR'' to "UVR" in the section title, above.}
\label{sec:baseline}

In this section, we evaluated the performance, particularly the visual
latency between the host PC and the MUD. of system frameworks that can carry
VR content untetheredly without relying on proprietary
hardware components. Our evaluation shows that all existing system frameworks
fail to meet the desired 20~ms deadline in avoiding motion sickness. We
further investigated the latency in each component of the fully open-sourced Gaming\-Anywhere 
platform~\cite{GamingAnywhere} and figured out the performance bottleneck in
existing systems. 

\subsection{Experimental methodology}
\label{sec:methodology}
This section describes the hardware platforms that we used to 
build the host PC and the MUD for selected frameworks as well as the
proposed \OpenUVR{} that this paper
will introduce later. We also explain how we measured
latency in all evaluated systems. 
%\CMFSRcom{It's unclear whether the first three paragraphs of section \ref{sec:hardware} 
%and all of section \ref{sec:measure} relate to \BaseUVR{} or \OpenUVR{} (or both) because neither system is explicitly mentioned in the text. My best guess is that you're discussing both \BaseUVR{} and \OpenUVR{}. If not, please revise accordingly.}

\subsubsection{Evaluated system platforms}
\label{sec:platforms}
We evaluated three platforms that can generate VR content on a desktop gaming PC
and communicate with an MUD through wireless links. We selected these platforms 
as they (1) rely on modern, commercialized wireless
technologies, (2) all support Unreal engine to enable fair
comparisons , and (3) at least allow the server software installed on
a custom-built gaming desktop.

These three platforms
are (1) NVIDIA's Moonlight~\cite{moonlight} that can use either a laptop or
a RPi 4b as MUD with only the client software open-sourced. (2) SteamLink~\cite{SteamLink} that uses a dedicated
wireless end-user device with an open-accessed API. (3)
GamingAnywhere~\cite{GamingAnywhere} that can use either a laptop or RPi 4b
as MUD, but fully open-sourced on both desktop PC server and MUD clients. In
addition to the above three, we also evaluated 
(4) \BaseUVR{}, an improved version of GamingAnywhere that we modified to
leverage NVIDIA's hardware accelerated NVENC on the host PC/laptop. On the
RPi4b MUD, \BaseUVR{} uses the OpenMAX Bellagio IL on~\cite{OpenMAX}
to write an H.264 packet into GPU memory and use the on-chip VideoCore 
IV accelerator to decode and directly render frames to the screen.

\subsubsection{Experimental platform/target hardware architecture}
\label{sec:hardware}
In all evaluated UVR architectures, the host PC uses a middle-class CPU and a
high-performance GPU, an assembly similar to most gaming desktop computers, to generate VR content.
The MUD of \BaseUVR{} and \OpenUVR{}, by contrast, consists of only a Raspberry Pi
computer with a wireless network-interface card (NIC). The wireless network uses IEEE 802.11ac
technology. When we tested each UVR framework, we found that some interference from other wireless 
networks was unavoidable; each use case operated within an environment that included 12 competing 
access points.

\paragraph{The host PC}
We built a host PC with a quad-core Intel processor based on the 
Coffee Lake architecture. The processor runs at a maximum clock rate of 
4.6~GHz and dynamically adjusts the voltage between 0.8~GHz and 4.6~GHz. 
The system has 16~GB DDR4 DRAM for main memory and
an NVIDIA RTX 2080 GPU attached to the host machine via 16\x{} PCIe Gen.3 lanes.
The NVIDIA RTX 2080 GPU provides hardware-accelerated NVENC encoder that
evaluted UVR frameworks can leverage. 
To provide wireless network capabilities, we installed a PCIe wireless card
with a Qualcomm Atheros chip and a 2\x{}5dBi omni-directional, external, detachable,
dual-band antenna supporting the IEEE 802.11ac standard at 867~Mbps.
\ignore{
To encode VR content into H.264 video format, the software of \BaseUVR{}/\OpenUVR{} 
on the host PC leverages FFmpeg's libavcodec library~\cite{ffmpeg}. The libavcodec library
allows the system to take advantage of the NVENC encoder. The host computer runs Ubuntu Linux 
(kernel version 4.8). }

\cfigure[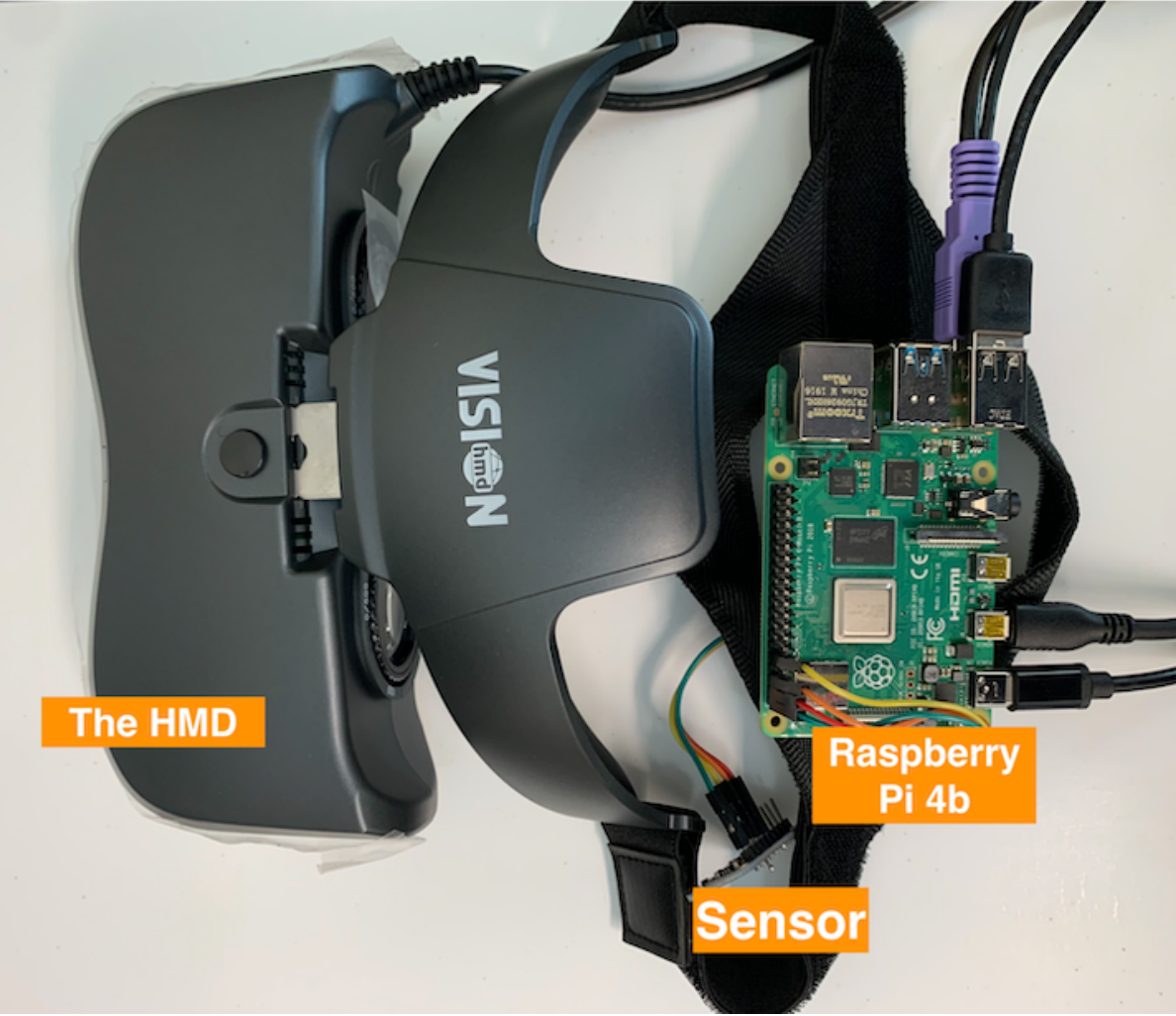, {The prototype MUD},fig:MUD]

\paragraph{The MUD} 
The target MUD architecture in \OpenUVR{} uses a Raspberry Pi
4 Model B (RPi4b) single-board computer that
costs only USD 35~\cite{RPi4}. RPi4b runs a simplified Linux distribution (Raspbian) on a
Broadcom BCM2711. The SoC runs at 1.5~GHz with built-in GPU cores and has hardware 
accelerators for decoding H.264 video/audio.
The RPi4b wireless NIC supports the IEEE 802.11ac
standard. The MUD can output VR content to a wearable display like the VISIONHMD Bigeyes H1 584PPI 
2.5K Screen 3D Video Glasses we used to test \BaseUVR{} and \OpenUVR{}. 
\ignore{
To decode VR content, the MUD leverages the OpenMAX Bellagio IL
library~\cite{OpenMAX}. This library allows
the software on the MUD to write an H.264 packet into GPU memory and use the on-chip VideoCore 
IV accelerator to decode and directly render frames to the screen.}
We also
installed gyroscopes and accelerometers on the MUD and wrote our own
programs to translate sensed signals into meaningful movements and
operations inside games. Figure~\ref{fig:MUD} shows the resulting MUD prototype. 

Although this work targets at system frameworks using ultra-light-weight MUD 
based on small and simple devices like RPi 4bs, we also tested the
performance on a Dell Inspiron 15 7559 laptop that uses an Intel Quad Core i7-6700HQ
processor with a dedicated NVIDIA Geforce GTX 960M GPU as a reference design
on MUD with intensive computing power.

\subsubsection{Performance measurements}
\label{sec:measure}
\cfigure[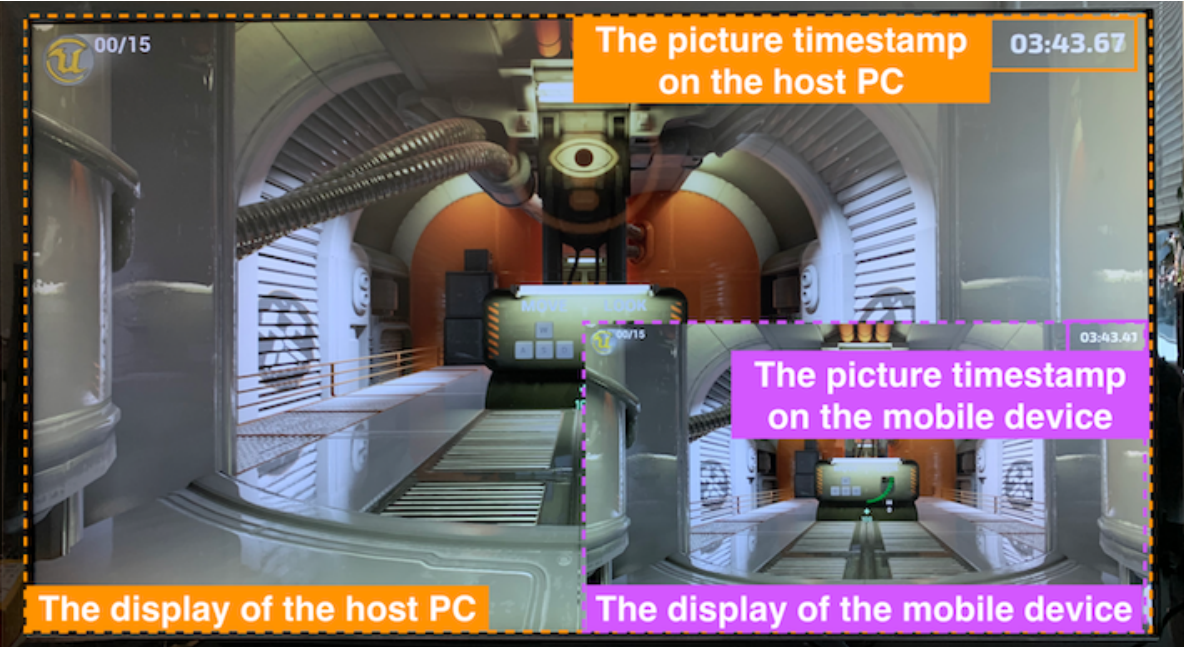, {A side-by-side
comparison of terminal output from the
host PC (orange) and the MUD (pink)},fig:exp_latency]

In evaluating the performance of VR systems, the most important metric is the
end-to-end latency between VR-content generation and VR-content
presentation. Unfortunately, fine-grain synchronization (e.g., in ~\us{}) of
clocks on different machines is impossible, so simply comparing the received
packet timestamp with the time on a different machine is not a feasible solution.

We addressed the difficulty of measuring
the latency between the host PC and the MUD by using photography (Figure~\ref{fig:exp_latency}).
We first turned on the timestamp feature in each game so that each rendered image
would contain a timestamp when generated on the host PC. 
We then displayed each image on a monitor connected to the host PC and
a separate monitor connected to the MUD, with the displays from
the two devices presented side-by-side. By using high-speed
photography to compare timestamps shown on each display, we were able to calculate
the latency between different nodes. As shown in
Figure~\ref{fig:exp_latency}, the host PC's timestamp is 03:43.67, but the
timestamp from the MUD is 03:43.41, reflecting a latency of 260~ms.
These photos also helped us assess the quality of delivered picture frames; for each framework 
being tested, we randomly sampled 1,000 frames. 

%Unless otherwise stated, all time measurements
%reported in this paper with
%units of \CMFdel{"}\CMF{``}frames\CMFdel{"}\CMF{''} are in terms of 90 \CMFdel{fps}\CMF{FPS}\CMFdel{ frames,
%unless otherwise stated. O}\CMF{, where o}ne 90\CMF{-FPS}\CMFdel{ fps} frame is
%equivalent to approximately 11.1~ms \CMFcom{Is there research to support this approach?}.
%Our VR content was rendered natively at 90 FPS on the host PC, each delayed frame

\ignore{
(as a plug-in module)
So we developed an instrumented version of Quake 3 Arena~\cite{Quake}, a plug-in for Unreal Engine
4, and a plug-in Unity Engine. Both Unreal and Unity Engines are widely used
as the Engines to power applications on modern VR systems. 
We recorded similar performance on these three instrumented games.}

By default, we tested the performance of these frameworks using
Unreal Tournament~\cite{Unreal}, an open-source, first-person shooter game.
Though Unreal Tournament is not, strictly speaking, designed for VR, the game is by far 
the best choice for testing a VR system for the following reasons:
(1) Unreal Tournament's first-person shooter gameplay allows the user to easily
perceive visual delays. (2) Unreal Tournament runs on an open-source game
engine (Unreal Engine), which makes optimization and code instrumentation
easy.
(3) Many popular VR systems like the Oculus and HTC Vive use Unreal Engine. 

To guarantee that motion was encoded when performance was measured, we used scripts that
emulate player input by causing a character to spin
counterclockwise at a constant rate while standing in place and keep moving
around in the scene for 10 minutes.
This virtual action effectively simulates the user behavior commonly found in VR applications. 

%the computation power
% of the \CMF{\OpenWVR{} }host PC.

\ignore{
To test the user-experience for immersive VR, we output the VR content from RPi3B+
to a VUZIX iWear Video Headphone that supports 720p displays for both eyes.
However, when quantifying the performance of 1080p resolution for
\OpenWVR{}, we output the motion picture frames to two displays that
support 1080p resolution.
}

\ignore{
In this section, we describe a baseline system implemented to investigate the deficiencies in, and potential 
improvements for, modern UVR systems that aim at latency-sensitive, real-time applications.
The system, called \BaseUVR{}, extends the open-source Gaming\-Anywhere platform~\cite{GamingAnywhere} to better
utilize hardware accelerators on both the host PC and the mobile device. In
this way, \BaseUVR{} significantly reduces Gaming\-Anywhere's demand on the MUD
device. The resulting \BaseUVR{} delivers competitive performance as
existing VR platforms that require more expensive MUDs, a justification for
\BaseUVR{} being a reasonable baseline for this paper.
We describe the datapath for \BaseUVR{}, the experimental
setup for \BaseUVR{}, and the measured latency for \BaseUVR{}. 
}
\subsection{The Performance of Existing Systems}
\label{sec:latencies}
\cfigure[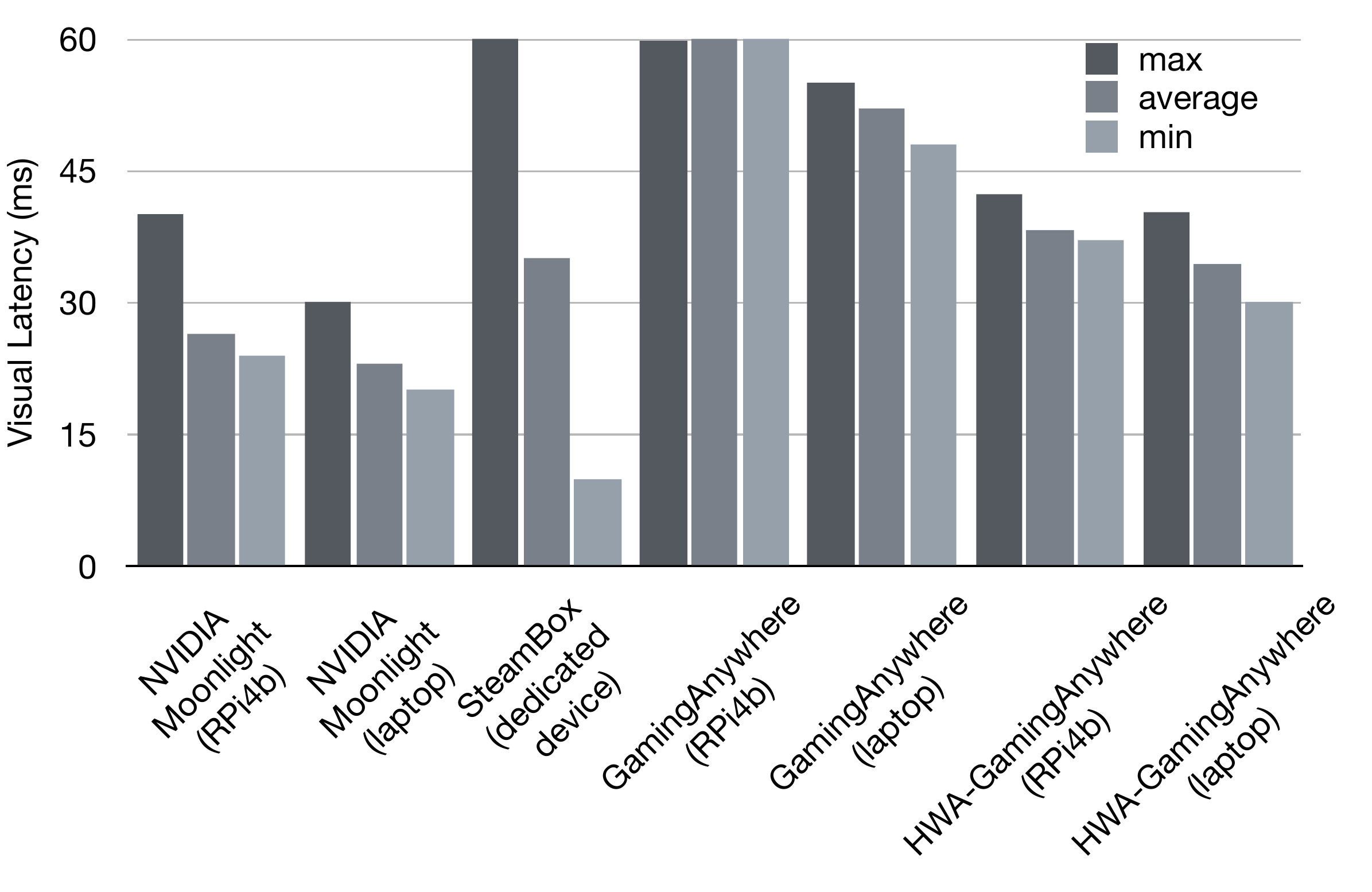, {Visual latency of evaluated
platforms},fig:base_latency]
\ignore{
 We evaluated three platforms that can generate VR content on a desktop gaming PC
and communicate with the MUD through wireless links. These three platforms
are (1) NVIDIA's Moonlight~\cite{moonlight} that can use either a laptop or
a RPi 4b as MUD with only the client software open-sourced. (2) SteamLink~\cite{SteamLink} that uses a dedicated
wireless end-user device with an open-accessed API. (3)
GamingAnywhere~\cite{GamingAnywhere} that can use either a laptop or RPi 4b
as MUD, but fully open-sourced on both desktop PC server and MUD clients. 
(4) \BaseUVR{}, an improved version of GamingAnywhere that we modified to
leverage NVIDIA's hardware accelerated NVENC. 
}

% \BaseUVR{}, compared with other commercialized game-streaming and VR platforms. 
Figure~\ref{fig:base_latency} shows the visual latency of evaluated
platforms. In general, platforms using laptops performs significantly better
visual latency than using RPi 4bs due to the more powerful computing power
on laptops. The original GamingAnywhere uses software video codec and the
latency is 1.96\x{} longer than Moonlight (i.e., 52 ms v.s. 26.5 ms) with the same client laptop. With
the help of NVENC, \BaseUVR{} delivers almost similar visual latency as
Moonlight, at 28~ms on average, when using a laptop as the MUD. 
Steamlink's dedicated device sometimes outperforms the visual
latency of using laptops. but still longer than Moonlight with a laptop MUD
at 35~ms on average. 
When Moonlight and \BaseUVR{} use RPi 4bs as MUDs, the visual latency of
both platforms
increases compared with using laptops. Moonlight achieves an average of
36~ms and \BaseUVR{} achieves around 38.3~ms. Without NVENC's hardware
acceleration, GamingAnywhere's latency on RPi 4b is as high as 1.2 sec.
However, no matter using what type of MUD, all platforms' average
visual latency failed the recommended 20~ms threshold to prevent 
motion sickness in real-time gaming. 

%of \BaseUVR{} is at the same level as those of the other
%platforms;
\subsection{The Sources of Latency}
\label{sec:base_performance}
\wfigure[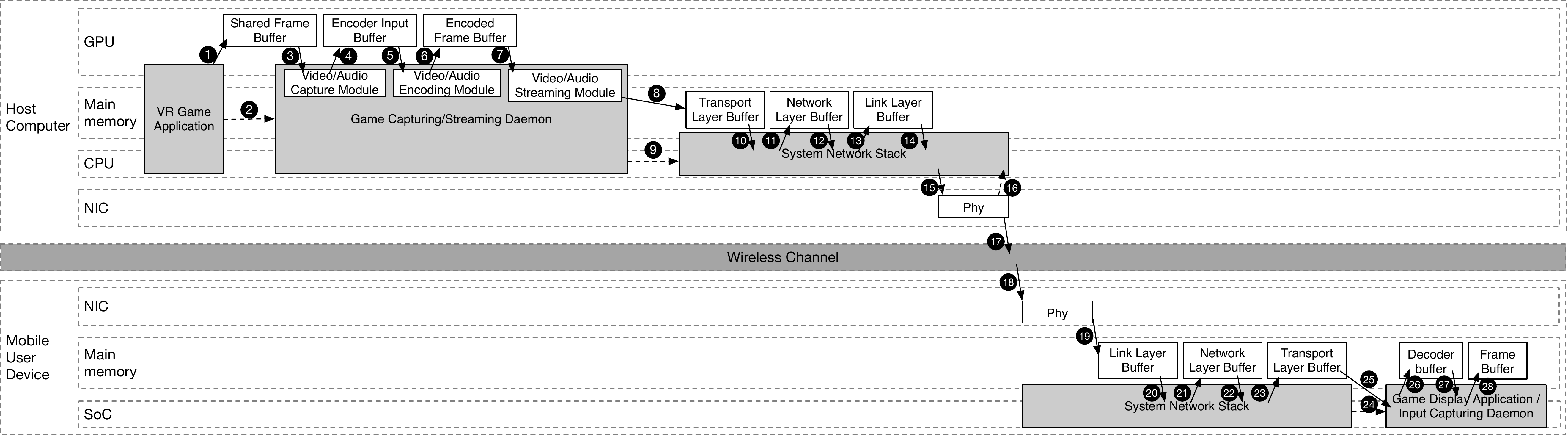, {The baseline architecture in a modern
UVR system},fig:baseline]

To figure out the sources of failed end-to-end latency for real-time VR
applications, we instrumented code of \BaseUVR{} system in (a) the host game application, 
(b) the game engine and the underlying library functions and system 
stack modules, and (c) the software stack on the MUD.
We only investigated \BaseUVR{} in deep and believe that a
detailed analysis of the latency in \BaseUVR{} is valuable for designing UVR systems 
that lack laptop-class computing power on the mobile user side
as (1) none of the platforms have reference designs that 
are fully open-sourced, (2) \BaseUVR{} delivers the same-level
performance as other counterparts, and (3) we investigated the Moonlight's MUD
code and the system implementation resembles \BaseUVR{}, showing
\BaseUVR{}'s system architecture is representative. 

\ignore{ 
To determine the sources of latency in \BaseUVR{}, we 
instrumented code in either (a) the host game application
or (b) the game engine and the underlying library
functions and system stack modules.
It's worth noting that \BaseUVR{} has only a minimally capable MUD. By contrast, some platforms 
(e.g., NVIDIA Moonlight and Gaming\-Anywhere) rely on systems such as the Dell Inspiron 15 7559 
laptop, while other platforms (e.g., SteamBox and Oculus) rely on dedicated closed-architecture
devices to achieve their reported
performance. The Oculus performance reported here enabled the V-Sync mode to
provide better visual quality. Without V-Sync mode, the latency is simply
5~ms. 
%\CMFSRcom{I've written "SteamBox" as it appears in the figures; however, "Steam Box" seems to be 
%the preferred form (two words, not one)}. 
If we port Gaming\-Anywhere directly
onto the RPi4b, the resulting visual latency is 1.2 sec on average, and the
video quality is low due to the huge processing overhead on the \BaseUVR{} MUD.
With the exception of Gaming\-Anywhere, none of the platforms have reference designs that 
are fully open-sourced, so a
detailed analysis of the latency in \BaseUVR{} is valuable for designing UVR systems 
that lack laptop-class computing power on the mobile user side. 

}

\subsubsection{Datapath of a frame in \BaseUVR{}}
Figure~\ref{fig:baseline} illustrates the datapath of a picture frame
and the frame's audio streaming from the host PC to the mobile device. UVR Frameworks
with a host PC that Section~\ref{sec:background_VR_systems}
describes~\cite{Furion,MoVR,ZhongWiGigGaming,CuttingtheCord,GamingAnywhere}
all use the same datapath. (Note that circled numbers in subsequent paragraphs correspond to the steps in Figure~\ref{fig:baseline}.)

When a VR application generates a picture frame \circled{1}, the
application signals \circled{2} a gaming
capturing/streaming daemon (usually a separate process that is allowed to access 
a shared frame buffer within the GPU) to retrieve the generated content
\circled{3}. To reduce the volume of encoded content, UVR
frameworks convert the captured content from RGB format into YUV color space
and place the converted content into an input buffer \circled{4}; this is 
typically done in GPU device memory to exploit the on-GPU hardware codec (e.g., 
NVIDIA's GPU-accelerated NVENC
video/audio encoder). 

Next, the UVR framework invokes the video/audio codec library to encode the 
transcoded content \circled{5} and places the result
into another buffer \circled{6}. Though modern GPUs support 4K
resolutions, we chose the default 1080p resolution that modern VR systems use, as prior work
suggests that 1080p resolution is sufficient for HMD~\cite{VRChallenges}. 
Further, our empirical study and work by Bao et al.~\cite{ShootingAMovingTarget} show that a 
20~Mbps bitrate effectively balances video/audio quality with bandwidth demand.

To deliver the encoded content to the destination mobile device, the UVR framework uses a
streaming library to feed content into the network stack's transport layer 
(\circled{7} and \circled{8}) and signals the network stack to take over \circled{9}. 
Data must then pass
through the network layer (\circled{\small{10}} and \circled{\small{11}}) and the link 
layer  (\circled{\small{12}} and \circled{\small{13}})
before reaching the network-interface hardware (steps \circled{\small{14}} through \circled{\small{17}}). 
Each network layer attaches its own protocol-specific header and checksum and
repartitions the data into the layer's own protocol-specific format and packet size. For example, UDP specifies a 
65,507-byte packet size, IPv4 specifies a 2,480-byte packet size, and IEEE 802.11ac specifies a
2,304-byte packet size. Consequently, CPU processing and multiple main-memory buffers 
are needed between steps \circled{8} and \circled{\small{15}}. 

Once the network-interface hardware on the mobile device has captured the
transmitted content \circled{\small{18}}, data move back up the network 
stack to the transport layer (\circled{\small{19}} through \circled{\small{23}}). 
Content then moves from the transport-layer buffer to the user-end application, and the 
user-end daemon receives a signal to take over processing
(\circled{\small{19}}---\circled{\small{24}} and \circled{\small{25}}). 
The user-end daemon uses its own computing resource (typically a system-on-chip [SoC] with 
combined CPU and GPU cores) to decode the content
(\circled{\small{26}} and \circled{\small{27}}). The daemon then hands off the decoded framework and audio
\circled{\small{28}} to an HMD-accessible buffer \circled{\small{28}}.

To improve the throughput and utilization of system components, modern UVR
systems like our own exploit pipeline parallelism (Figure~\ref{fig:baseline}). But
pipelining does not eliminate the need for each frame to go through every step in the 
datapath outlined above. Rather, pipelining increases latency due to the overhead of synchronizing pipeline stages.

\subsubsection{Latency analysis of \BaseUVR{}}
\cfigure[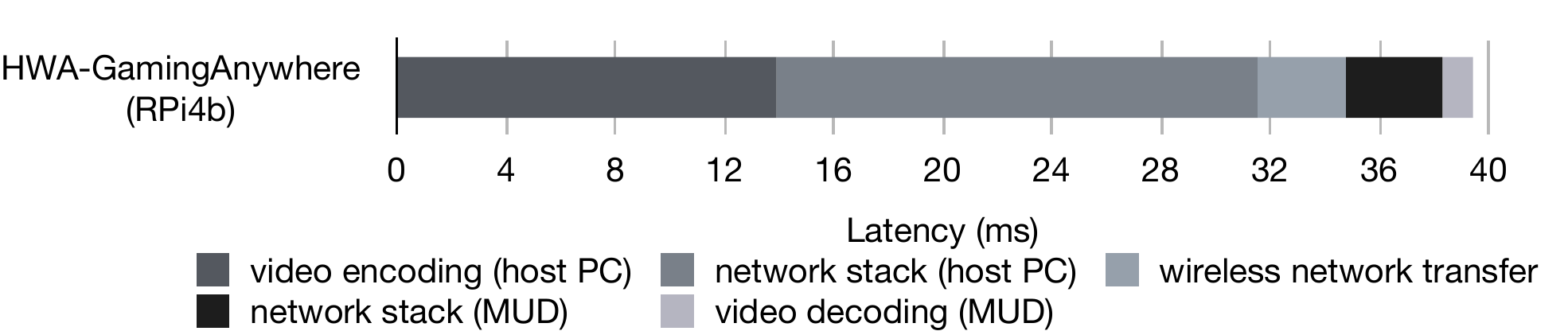, {Latency breakdown in \BaseUVR{}},fig:breakdown]

Figure~\ref{fig:breakdown} breaks down the latency in each stage of the VR
system pipeline. The most significant source of latency comes from the
network stack on the host PC, which includes the transport-layer protocol (i.e., RTP/RTSP),
the network-layer protocol (i.e., IPv4), and the link-layer soft MAC
firmware. The network stack on the host PC contributes 17.63~ms to the
end-to-end latency on our testbed. The video encoding latency, which contributes 13.94~ms to the total latency, 
ranks in second place. The aggregate latency measured in the host PC software stack is 31.57~ms. 
%Our observation suggests these two parts as the 

On the other hand, the fixed cost in the network subsystem is only 3.2~ms; this includes propagation delay, physical
modulation/demodulation, and link-layer control. The total latency on the MUD is only 3.64~ms, as \BaseUVR{} 
requires minimal computation from the MUD while utilizing accelerators on the highly optimized SoC. 

In summary, our measurements indicate that if we can minimize software
latency, we can reduce end-to-end latency and prevent motion
sickness in the user---and this can be done with existing wireless network technologies and a simple, low-cost
MUD.

\cfigure[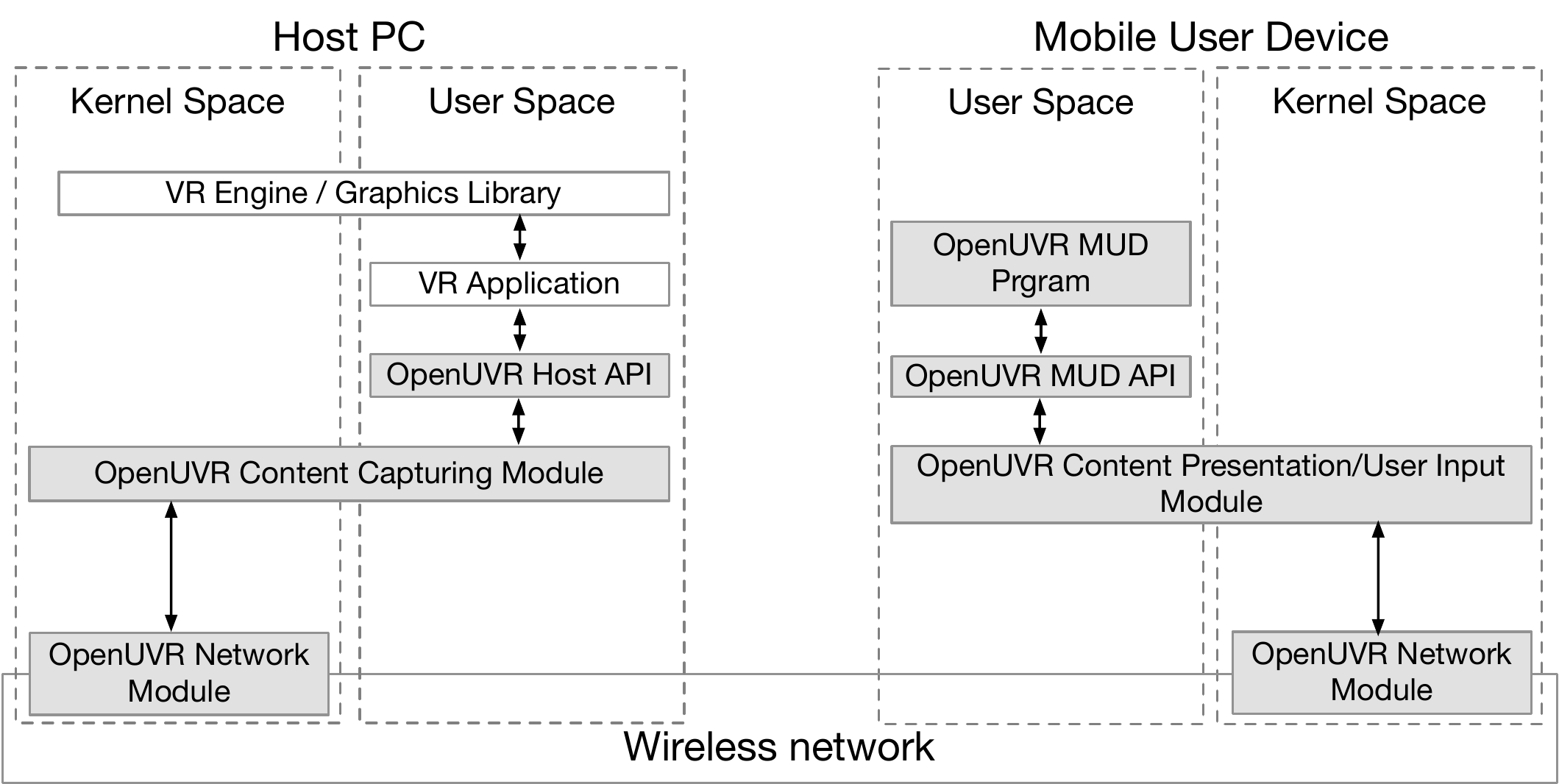, {The \OpenWVR{} system architecture \CMFSRdel{of
\OpenWVR{}}},fig:system_stack]

\section{\OPENUVR{} System Design}
\label{sec:arch}

With lessons learned from implementing \BaseUVR{}, we identified the
potential of fulfilling the 20~ms threshold that prevents motion sickness
through simply reducing datapath software latency. 
%By revisiting the
%datapath in Figure~\ref{fig:baseline}, we focus on 
In response to the potential points of optimizations, this paper presents the \OpenUVR{}
system.
Figure~\ref{fig:system_stack} shows the system stack of \OpenUVR{}. On the
host-PC side, \OpenUVR{} receives data from an application through the \OpenUVR{} API,
uses the \OpenUVR{} content-capturing module to encode VR content, and
delivers the encoded content to the MUD via the \OpenUVR{} network module. 
Once the network module has receives data on the MUD side, the \OpenUVR{}
content-presentation module will decodes the content and allows the user-space MUD program 
to display the decoded data to the user. If the user experiences any feedback through sensors 
or a joystick, the MUD program captures the input and sends it to the host PC in reverse order. 

\wfigure[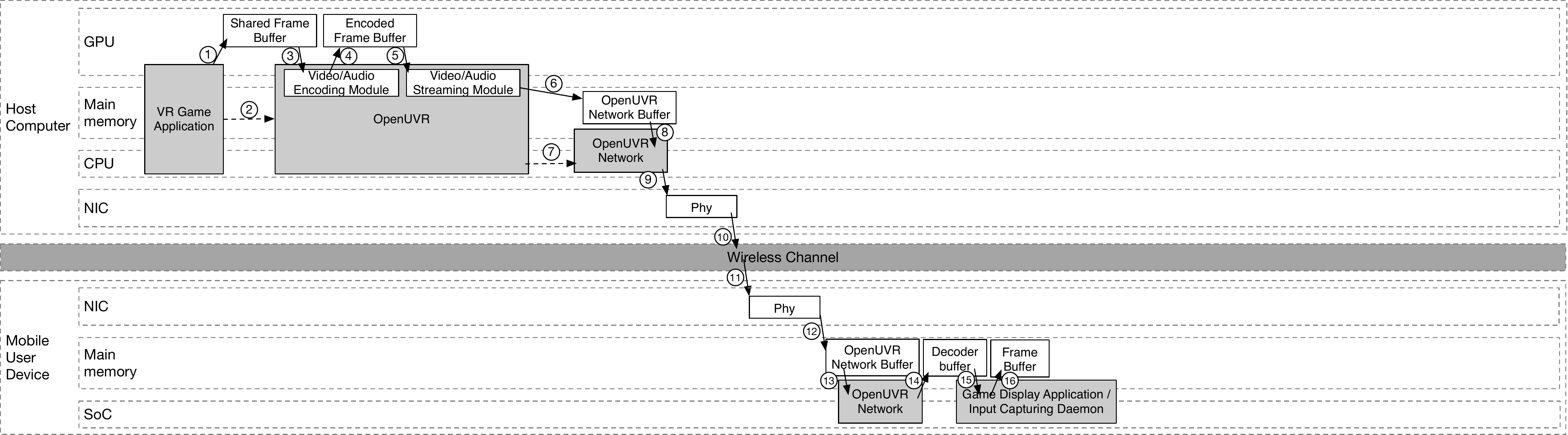, {The \OpenWVR{} datapath\CMFSRdel{ of
\OpenWVR{}}},fig:overview]
In designing each \OpenUVR{} module, we carefully optimized the software
latency to eliminate (1) unnecessary data transcoding, (2) unnecessary memory copies, and
(3) system-level control overhead. 
The system-stack architecture allows each framework to flow through the
datapath shown in Figure~\ref{fig:overview}---a datapath that is significantly more streamlined
than the datapath in Figure~\ref{fig:baseline}. 
The following sections describe our design and the techniques we used to reduce latency. 

\subsection{\OpenUVR{} content-capturing module}
\label{sec:video}
Because the \OpenWVR{} content-capturing module runs on the host
PC, the module can intercept VR-application content while working closely with 
real-time video/audio-compression hardware accelerators before delivering 
the encoded content to the network module. 
The \OpenUVR{} content-capturing module
includes three sources of optimization to minimize unnecessary operations and
reduce latency: transcoding avoidance, plug-in modules, and kernel-space
functions. 

\subsubsection{Transcoding avoidance}
In \BaseUVR{}, the largest contributor to latency in the game-streaming
daemon arises from converting the GPU-generated VR content from RGB format to YUV format. 
(YUV format is generally preferred for video-compression applications because it allows 
efficient lossy compression to reduce video size without having an appreciable effect on 
perceived video quality.) \BaseUVR{}, Furion and Gaming\-Anywhere use YUV420, which splits 
an image into its components of luma  (black and white intensity) and 2-dimensional chrominance 
(color). YUV420 downsamples both  chrominance components in the horizontal and vertical dimensions, 
thereby reducing the data required for each component by 50\%. 

In spite of the advantages that YUV420 offers, conversion between two different color spaces can 
incur considerable overhead. YUV420 may therefore contribute to latency between content generation and
content perception. 
Additionally, color-space conversion not only consumes computing resources and time, but also requires 
memory space and memory access to accommodate
the raw RGB content and the converted YUV content (steps \circled{4} and \circled{5} in Figure~\ref{fig:baseline}). 
In conventional video-streaming applications, the added latency from
color-space conversion and buffering is acceptable because these applications can tolerate
longer delays or do not require a user to interact with the content generator. 
Such applications compensate for the delays from color-space conversions by relying on 
higher video quality and lower bandwidth usage as well.

As VR applications are extremely latency sensitive, \OpenWVR{}
avoids color-space conversions, and so also avoids color-space transcoding. \OpenWVR{} can 
leverages the hardware-accelerated encoder in the graphics card to compress VR content 
while the content is in native RGB format. \OpenWVR{} then relies on the hardware accelerator 
in the RPi4b MUD's SoC to decode the RGB data.
In this way, \OpenUVR{} is able to simplifies the
process of generating encoded content from steps \circled{3} through \circled{6} in
Figure~\ref{fig:baseline} to just two steps---steps \circlel{3} and \circlel{4} in
Figure~\ref{fig:overview}.

\subsubsection{Plug-in modules for VR applications}
Unlike conventional VR solutions that use a daemon to
capture application content, \OpenWVR{} requires
an application to either directly invoke \OpenWVR{}'s API
functions or indirectly invoke them through a plug-in module in the game
engine. The \OpenWVR{} API functions do not create additional processes
in the system. Rather, the functions all work within the same
process, so they share the address space of the calling application. In this
way, \OpenWVR{} avoids the context-switching overhead and memory copies
associated with inter-process communication and makes the signaling between
the VR application and \OpenUVR{} modules more efficient. 

Because \OpenUVR{} and the VR game share both
the host OS and GPU virtual memory space, 
the device memory locations on hardware accelerators are visible to
the \OpenUVR{} content-capturing module. \OpenUVR{}
takes advantage of this visibility and avoids redundant memory copies by
integrating hardware-accelerated codecs.
Since a VR application can associate a GPU-device
memory buffer with an \OpenWVR{} context (which allows the application to share
the same memory space as the context), \OpenWVR{} can poll the GPU memory buffer 
periodically and invoke the video-encoding library on the GPU memory buffer directly; \OpenWVR{}
simply needs to move the final encoded VR content into the main memory minus the raw VR content 
that the GPU generates. Without this optimization, the current GPU runtime would need
to either explicitly copy raw rendered data from the GPU memory to a shared memory location
(e.g., texture memory) or use inter-process communication
facilities (most likely the host DRAM).

\subsubsection{Kernel-space functions}
The \OpenWVR{} content-capturing/presentation modules implement some of
their functions as system kernel modules.
For instance, after the
API has initialized the \OpenWVR{} context, the
content-capturing module acts as a kernel module when it remaps
the user-space main-memory locations to the same physical memory locations
as those accessed by the system wireless-network
stack (e.g., the NIC driver). This design allows
the application data to directly reach the network stack without
additional copies from user-space to kernel-space (steps \circlel{6} and \circlel{14} in Figure~\ref{fig:overview}).

\subsection{Wireless network subsystem}
To tackle the most significant latency in the VR datapath, the wireless
network subsystem in \OpenUVR{} applies several design decisions that
directly or indirectly lead to latency reduction. 

%\subsubsection{\OpenWVR{} network stack/protocols}
%\ignore{
\subsubsection{Direct peer-to-peer network topology}
\OpenWVR{} adopts the peer-to-peer (P2P) network topology
specified in IEEE 802.11ac---a topology that
allows the host computer and MUD to directly communicate with each other
(unlike the conventional infrastructure mode that
relies on a base station/router as an intermediary for wireless communication).

Using P2P mode provides three benefits.
First, the mode reduces the latency on wireless channels
by eliminating the one-hop delay on the wireless base station. Though this
latency is only 3.2~ms in \BaseUVR{}, the latency can double as
\OpenUVR{} generates RGB encoded data that can increase the data size of each frame. 
Second, P2P mode halves congestion on wireless channels because the host PC and the MUD 
usually share the same
access point in \BaseUVR{} and \OpenUVR{}; using P2P mode eliminates the need for a 
datagram to go through the wireless channels within the same area twice. Third,
direct communication between node pairs using dedicated channels removes the demands of the
network-layer protocol; this, in turn, gets rid of memory
copies from packet reformatting (steps \circled{10} and \circled{11} in
Figure~\ref{fig:baseline}).
In the current \OpenWVR{} implementation, peers are connected on an 80-MHz-wide channel 
within the 5~GHz spectrum, and hostapd is configured to select the channel 
on which it detects the least interference from other radio broadcasts.

\subsubsection{Direct network I/O---\OpenUVR{}DPP}
The \OpenWVR{} data plane protocol (\OpenWVR{}DPP) is
customized to eliminate memory copies when data packets are reformed and headers attached in each
layer of data-plane traffic. As real-time gaming is especially latency
sensitive, the best transport-layer protocol for data traffic should neither
perform additional operations nor use additional buffers (essentially a UDP protocol with
no-ops). P2P network topology removes the need for the
network-layer protocol and for packet reformation, allowing \OpenUVR{} to bypass these two 
layers and minimize latency. 

So instead of using the port number abstraction of UDP/RTP,
\OpenUVR{}DPP works with the \OpenUVR{} kernel module to map the physical memory location of
the data-link-layer buffer to the VR application's memory
space. Given such access, \OpenUVR{} can copy GPU data
directly into \OpenUVR{}DPP's data-link-layer buffer (step \circlel{8} in
Figure~\ref{fig:overview}) and generate packets without going through any conventional
routing- and transport-layer protocols.

\subsubsection{Connection/quality control---\OpenWVR{}CP}
To set up an \OpenUVR{} connection and provide basic quality control, \OpenUVR{} uses an 
off-band control protocol (\OpenWVR{}CP). \OpenWVR{}CP is relatively less time 
critical and less bandwidth-consuming than \OpenWVR{}DPP. Thus, we did not implement a transport-layer bypass. Instead, we adopted
UDP to carry \OpenWVR{}CP control messages.

\paragraph{Connection setup}
When a MUD launches its instances and grants access to the host PC's P2P network, the MUD sends
an \OpenWVR{}CP message to the host PC to request access. If the user chooses to grant the MUD's 
access request, the user adds the MUD's MAC address to the whitelist or goes through the host 
PC's administrative user interface. 

\paragraph{Quality control}
As \OpenWVR{} still relies on video compression to reduce bandwidth
demand, video-codec parameters play an important role in the
trade-offs between latency, visual quality, and network bandwidth. Among the
most important parameters is the size of a group of pictures (GOP).
In the H.264 format that \OpenWVR{} currently adopts, a GOP consists of an I-frame followed by
a certain number of P-frames, where an I-frame represents the encoding of a
standalone image and a P-frame encodes only the changes between the current
frame and the previous frame. (The H.264 standard also defines a B-frame, which we
don't use since it relies on the frame that comes after it, making a B-frame unsuitable
for live-streaming applications.) The GOP size defines the number of
P-frames between each I-frame. P-frames are smaller than I-frames (about a quarter
the size of an I-frame), so at a given constant target bitrate, having a larger
GOP size should result in smaller data size; however, if the client drops
a frame, the display will be corrupted for all following P-frames because each frame
builds on the content of a previous frame (and a dropped frame never 
arrives). This dependency means that streams with larger GOP sizes will, on average, 
be corrupted for a longer period of time when a frame is dropped.

In general, a larger GOP size reduces bandwidth demand, but the larger size also
makes an encoded stream more vulnerable to network packet loss. A smaller GOP size
increases the size of data transmissions and makes the encoded data more
tolerant of packet loss. However, smaller GOP sizes increase encoding
latency and bandwidth usage. Smaller GOP sizes create higher demands on GPU-device
memory and system main memory.

In order to reduce latency while decreasing the disruption of visual stuttering, 
\OpenWVR{} allows the MUD to send feedback to the host PC regarding dropped frames.
When the MUD video decoder drops a frame because it hasn't received all of the data 
within a certain time limit, the MUD program triggers \OpenWVR{}CP to send a short 
message to the host PC requesting that the host PC generate an I-frame on the next transmission.
The MUD sends this control message after every frame until it receives an I-frame.
When the host fulfills an I-frame request, it ignores additional requests from the MUD for a 
predetermined period of time to prevent the generation of consecutive large frames that would 
worsen network conditions.

%It also allows us to decrease the
%target bandwidth to 20 Mbps, which decreases the network latency by about 0.5 ms, the

%\subsubsection{The network topology}
\subsection{\OpenWVR{} API}
\label{sec:API}
\begin{table*}
\centering
\scriptsize
%\footnotesize
\begin{tabular}{|p{3.6in}|p{3in}|}
\hline
Synopsis & Description \\
\hline
\multicolumn{2}{|l|}{\OpenWVR{} Host API}\\
\hline
\texttt{struct openuvr\_context
*openuvr\_alloc\_context\newline(enum OPENUVR\_DECODER\_TYPE dec\_type,
\newline
enum
* OPENUVR\_NETWORK\_TYPE net\_proto, \newline
void *ptr)} &
Initialize an \OpenWVR{} context with \CMF{the }specified video encoder, the desired
network protocol\CMF{,} and an OpenGL pixel buffer.
\\
\hline
\texttt{int openuvr\_set\_feature(struct openuvr\_context *context,\newline
enum OPENUVR\_FEATURE feature,
int value)} &
Configure a feature of an existing openuvr\_context with the specified value\CMFSR{.}
\\
\hline
\texttt{void openuvr\_init\_thread(struct openuvr\_context *context)} &
Create an \OpenWVR{} thread using the given \OpenWVR{} context.
\\
\hline
\texttt{int openuvr\_map\_input\_device(struct openuvr\_mud\_context\newline
*context, int id, int fd); }
&
Map a device registered with \texttt{id} on \CMF{the }\OpenWVR{} MUD device to a local
device using its opened descriptor.
\\
\hline
\multicolumn{2}{|l|}{\OpenWVR{} MUD API}\\
\hline
\texttt{struct openuvr\_mud\_context *openuvr\_alloc\_mud\_context(enum
OPENUVR\_DECODER\_TYPE dec\_type, \newline
enum OPENUVR\_NETWORK\_TYPE net\_proto);} &
Create an \OpenWVR{} MUD context with the specified decoder and \CMFdel{the }network
protocol.
\\
\hline
\texttt{int openuvr\_init\_mud\_thread(struct openuvr\_mud\_context\newline *context);
} &
Create a\CMFdel{n} MUD thread th\CMFdel{e}\CMF{at} \CMF{captures inputs and }renders the received VR content using \CMF{a }standard
display device\CMFdel{ and captures inputs}.
\\
\hline
\texttt{int openuvr\_register\_input\_device(struct openuvr\_mud\_context
*context, int id, int fd);
} &
Register an input device to a device \CMF{\texttt{id}}\CMFdel{id} in \OpenWVR{} using an opened descriptor.
\\
\hline
\end{tabular}
%\vspace{-1mm}
\caption{The \OpenWVR{} API\CMFSRdel{.}}
\label{table:API}
\vspace{-0.3in}
\end{table*}

%\cfigure[Figures/eud_code.pdf, {The initialization code in the VR MUD application to
%setup \OpenWVR{}},fig:code_eud]
\cfiguredouble[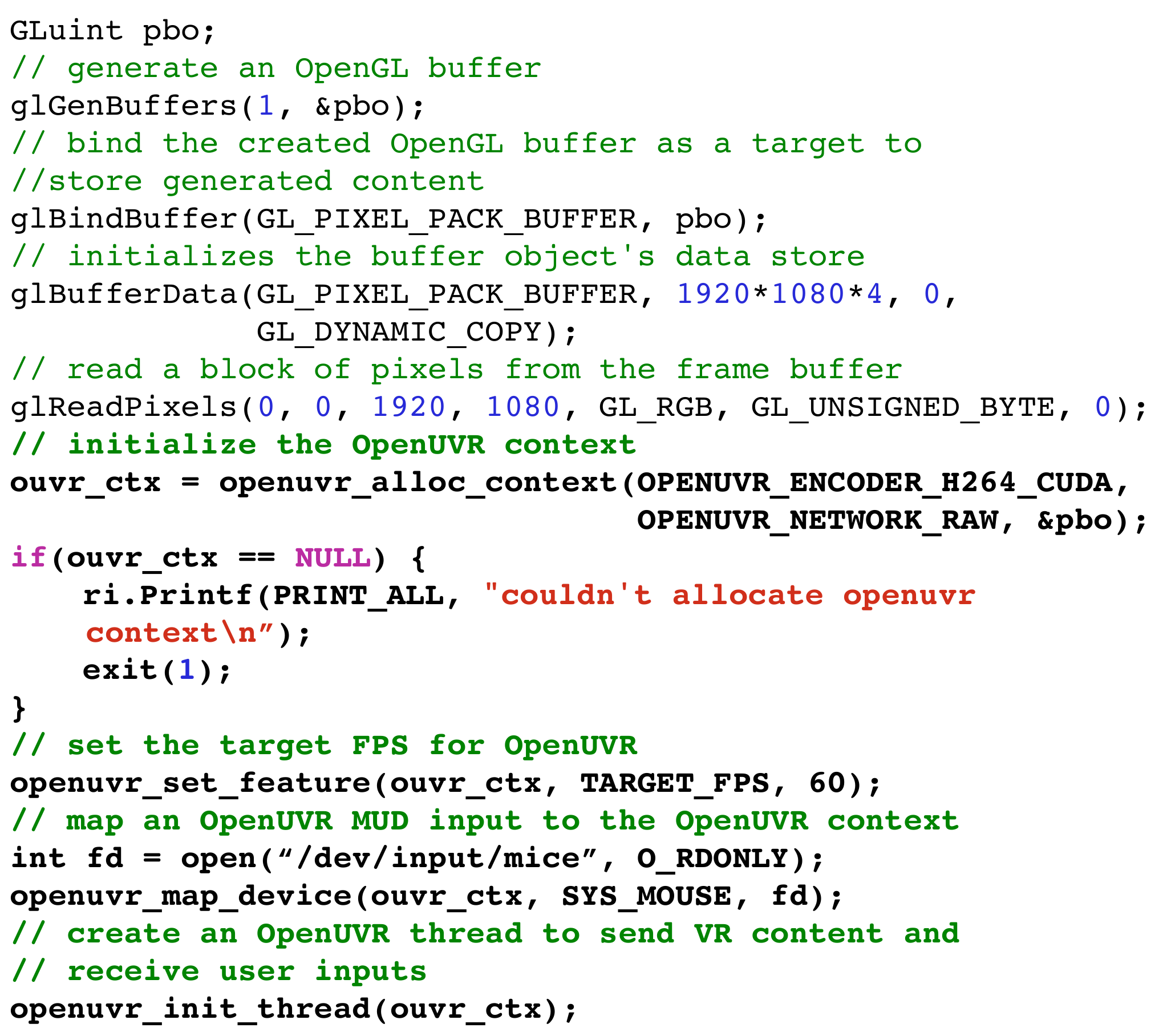,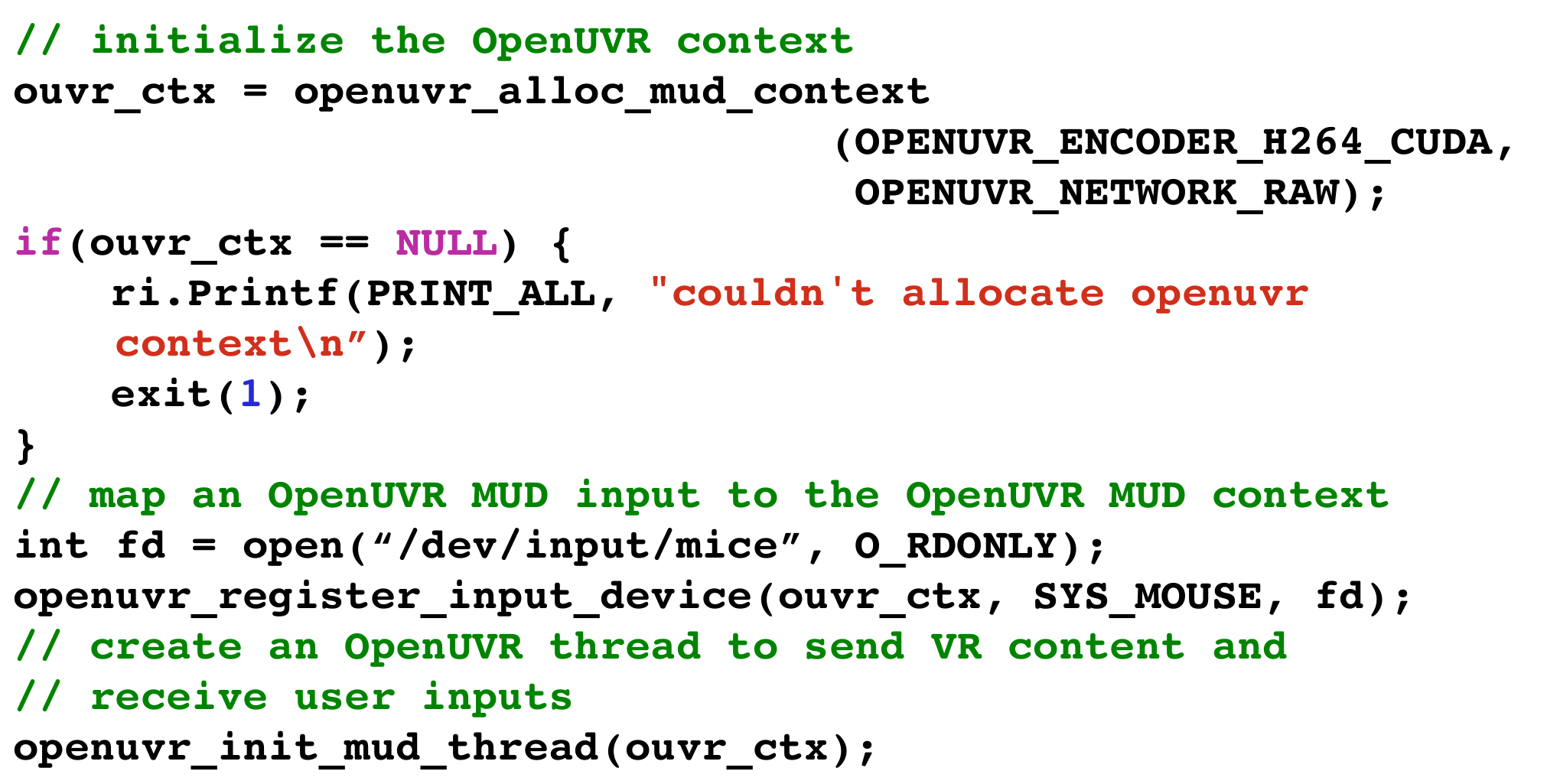, {The \OpenWVR{} setup-initialization code in
(a) the VR host application and (b) the MUD daemon},fig:code]

To enable \OpenWVR{} in a VR application on the host PC, a
programmer need only add a
few function calls in the \OpenWVR{} API\CMFdel{ to the application,} or
\CMFdel{implement}\CMF{have} a
plug-in module invoke these API functions \CMF{(}if \CMFdel{the application framework
supports}\CMF{supported by the application framework}\CMF{)}. \CMFdel{Similarly, t}\CMF{T}o collect inputs and
display VR content, the system \CMFdel{also }needs
\CMFdel{a}\CMF{the MUD} daemon\CMFdel{ program on the MUD that}\CMF{ to}
invoke\CMFdel{s} the \OpenWVR{} MUD API. Table~\ref{table:API} lists the
basic functions \CMF{necessary for the host PC and MUD }to initiate \OpenWVR{}
in an application. \CMFdel{ for both the host PC and the MUD} 
%\CMFSRcom{Please consider making the following changes to Table~\ref{table:API}: (1) add horizontal padding to each row, (2) remove some of the row and column borders, (3) move the ``OpenUVR Host API'' title row above the Synopsis/Description header row, and (4) duplicate the Synopsis/Description header row beneath the ``OpenUVR MUD API'' title row. In Figure~\ref{fig:code}, please change the comment ``// initializes the buffer object's data store'' to ``// initialize the buffer object's data store'' so the comment is consistent with the other comments.}

Figure~\ref{fig:code}(a) shows an example of code that uses
these API functions in a VR application on the host PC.
In the initialization
\CMFdel{part of a host application that generates the VR content}\CMF{section}, the\CMFdel{ program}
code\CMFdel{ needs first} create\CMF{s}\CMFdel{ a few} buffer objects for OpenGL
\CMF{(}or \CMF{an}other graphic\CMFdel{al}\CMF{s}
librar\CMFdel{ies}\CMF{y)} to store the generated content. \CMFdel{The
program invokes \OpenWVR{} functions a}\CMF{A}fter these buffer objects are
created\CMFdel{. T}\CMF{, t}he code
\CMF{uses the \texttt{openuvr\_alloc\_context} function to associate}\CMFdel{first creates}
 an \OpenWVR{} context\CMFdel{ that is assoicaciated} with a
buffer object\CMFdel{ using the \texttt{openuvr\_alloc\_context} function. This API}\CMF{; the}
function \CMF{also }allows the programmer to specify \CMFdel{the desired}\CMF{both a}
video codec and \CMFdel{the}\CMF{a} network protocol for
carrying the VR content. Once the program \CMF{has }successfully set\CMF{ }up an
\OpenWVR{} context, the program can\CMFdel{ then} set\CMFdel{ several} features for the
context using \texttt{openuvr\_set\_feature}\CMFdel{ and finally}\CMF{. The final portion of the code}
launch\CMF{es} a thread \CMF{that }run\CMFdel{ning}\CMF{s} concurrently with
the VR application to deliver\CMFdel{ the} content and
\CMF{handle }signal\CMFdel{s}\CMF{ing} between the host \CMF{PC }and the MUD.

Figure~\ref{fig:code}(b) shows code that the MUD daemon runs. \CMFdel{On the MUD, t}\CMF{T}he
daemon\CMFdel{ program needs to} first initiate\CMF{s}
an \texttt{openuvr\_mud\_context} using the \texttt{openuvr\_alloc\_mud\_context}
function. \CMFdel{T}\CMF{Next, t}he program\CMFdel{ will then} register\CMF{s} each input device\CMFdel{s/sensors} \CMF{(}using
the corresponding descriptor\CMF{)} and associate\CMF{s} \CMFdel{with a}\CMF{each}
device\CMF{ with an} ID \CMFdel{that}\CMF{known by} both the
MUD and the host-PC application\CMFdel{ knows}. The program\CMFdel{ will} then
invoke\CMF{s} the \texttt{openwvr\_init\_thread} function to create a thread that
\CMFdel{keeps displaying}\CMF{continually displays} the received VR content and
poll\CMFdel{ing the}\CMF{s for} user inputs.

\section{\OpenWVR{} performance}
\label{sec:result}
%\CMFSRcom{I changed ``PERFORMANCE OF OPENUVR'' to "OPENUVR PERFORMANCE" in the section title, above.}
%\input{tableLatency}
Using the observations from Section~\ref{sec:baseline},
we optimized the \OpenWVR{} system stacks to further reduce end-to-end latency.
We then used visual-latency performance assessments, a user study,
and differing hardware configurations to evaluate the optimizations.
%\OpenWVR{} proposes and integrates the following optimizations in various
%layers of the system to address the delay in the basic implementation.

\subsection{Latency}
\label{sec:latency}
In this section, we evaluate the visual latency of \OpenUVR{} 
and describe how each optimization helps reduce latency. 

\subsubsection{Visual Latency}
\cfigure[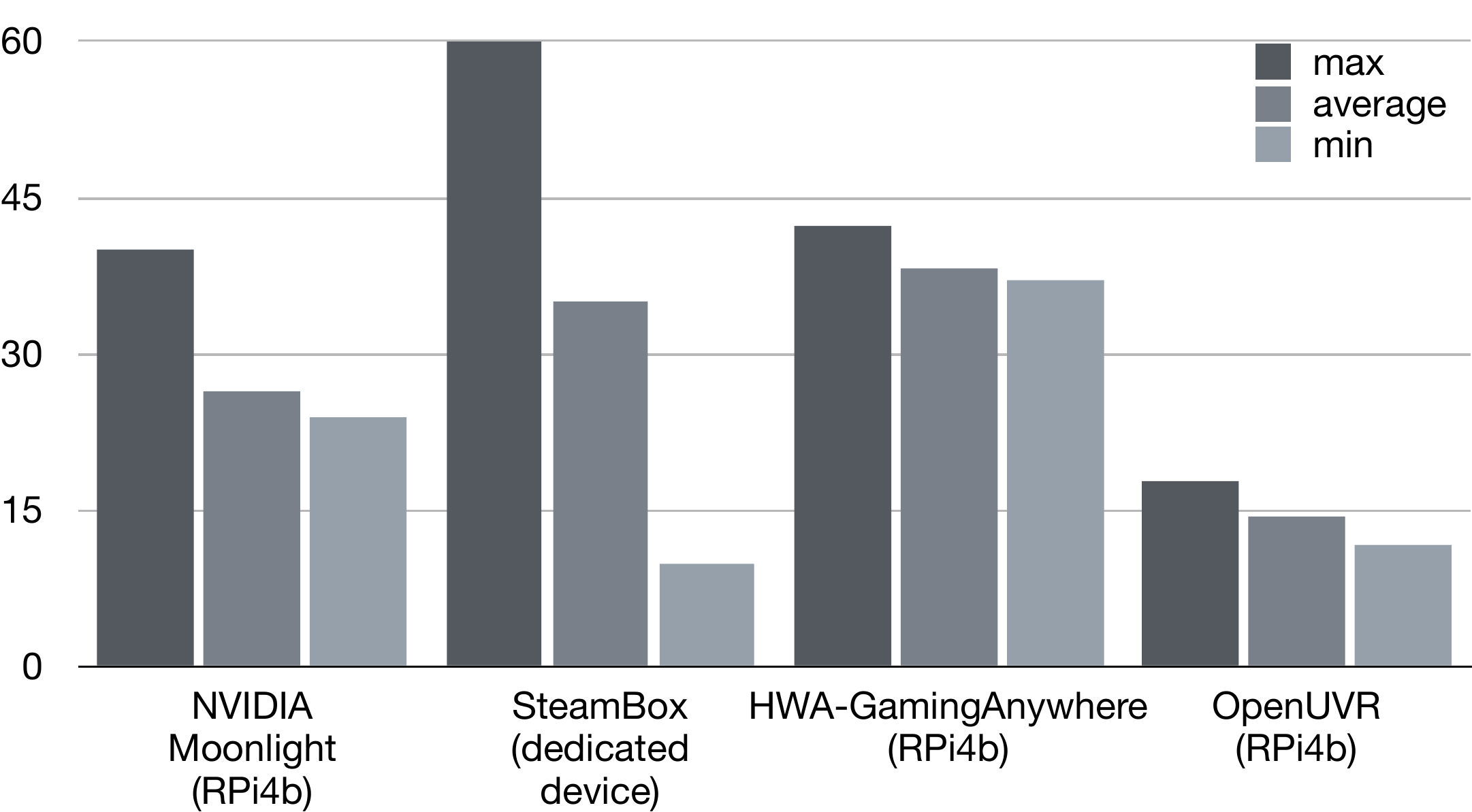, {The visual latencies of \OpenUVR{}, \BaseUVR{}, and other VR systems},fig:visual_latency]

Figure~\ref{fig:visual_latency} compares the visual latency of \OpenUVR{} with that of \BaseUVR{} and other VR systems such as the Oculus Rift.
The average latency of \OpenUVR{} is only 14.32~ms, which is 2.67\x{}
better than the average latency of \BaseUVR{}. \OpenUVR{}'s latency is also lower than the
latency of the 60-FPS display that an RPi4b can support. Of crucial importance is the fact 
that \OpenUVR{}'s latency falls below the threshold for motion-sickness onset. 

\subsubsection{Latency reductions from optimizations}
\label{sec:latency_opt}
\cfigure[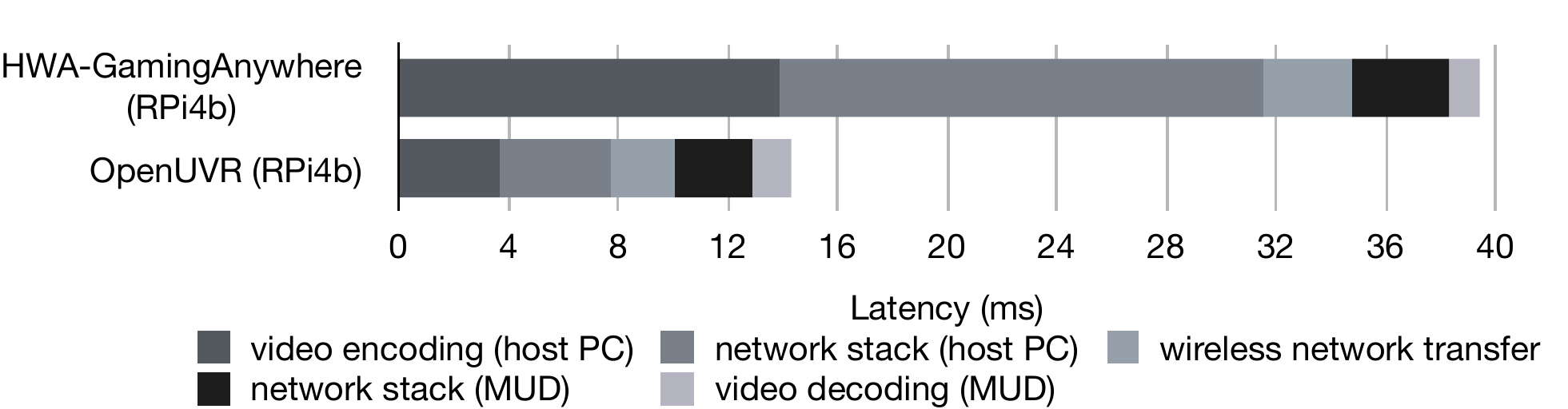, {The latency breakdown for \BaseUVR{} and
\OpenUVR{}},fig:latency_optimization]
Figure~\ref{fig:latency_optimization} summarizes the effect of each optimization
discussed in
Section~\ref{sec:arch} and shows how \OpenWVR{} brings the
average visual delay down to less than one 60-FPS frame.
The most significant reductions in latency were produced on the host PC, with
minor improvements observed for the MUD.

By avoiding the color-space conversion, \OpenUVR{} removes 5.51~ms of
end-to-end latency. Though using RGB encoding increases data size, the increase is offset by the 
P2P network topology, which
reduces network-transfer latency by 0.8~ms. As a reference, with P2P and YUV encoding
transfer latency reduced by 1.6~ms.

Data sharing between \OpenWVR{} and the running application reduces end-to-end latency by 4.71~ms. 
The data sharing is made possible through use of the \OpenWVR{} API for applications and the plug-in module 
for the underlying game engine). The direct-network I/O mechanism lowers network-stack latency by 13.67~ms on 
the host PC and 0.7~ms on the MUD. \OpenWVR{}CP's feedback control improves overall latency by only 0.1~ms, but 
the resulting video quality is far superior to the quality achieved without feedback control.

\subsection{Energy consumption}
We also evaluated the energy consumption of the \OpenUVR{} system. We used a single USB power bank 
with a maximum capacity of 5,400 mAh to simultaneously supply the
whole MUD, the RPi4b, and the HMD (with RPi4b output connected to
the HMD). This single-power-supply setup allowed the MUD to run for 6 hours and 20 minutes. As
a reference, a fully charged Oculus Quest running for 8 hours consumes 10,000 mAh
---significantly more energy than \OpenUVR{} needs~\cite{OculusQuestBattery}.

\subsection{Performance sensitivity of GPU architectures}
As the main computation resource for generating VR content and encoding
H.264 data, the GPU plays an important role in a VR system.
In addition to using the RTX 2080 GPU (mentioned in Section\ref{sec:methodology}) to analyze \OpenWVR{}'s performance
, we used the NVIDIA GTX 1060. The GTX 1060 has the smaller amount of memory
 than the RTX 2080 (6~GB v.s. 8GB) and uses
an earlier Pascal microarchitecture and runs at about the same frequencies
(1708~MHz vs. 1710~MHz).

Both cards can stably generate VR content at 90 FPS on the host side; however, the
GTX 1060 has an end-to-end latency that is 2~ms higher than that of the RTX 2080. 
\OpenUVR{} still fulfills the 20~ms threshold even with a mid-range
video card. 
The source of increased latency is the longer encoding time that the GTX 1060 requires.
The performance of hardware-accelerated video encoding for modern GPU architectures thus has 
a significant impact on end-to-end latency in the current version of \OpenWVR{}.

\subsection{Quality}
\label{sec:user_study}
To evaluate the quality of \OpenUVR{}, we conducted two types of
experiments. A quantitative analysis and a series of user study. 

\subsubsection{Quantitative analysis}
\label{sec:ssim}
To calculate the Structural Similarity (SSIM), a quantitative measurement
of loss in frame quality between the original version and the
user-perceived version that the video processing community uses~\cite{SSIM},
we instrumented code on both the \OpenUVR{} content capturing module and the MUD
moudle to record the received frames on both sides. 
The modified modules
will record 150 successive frames for every 10 seconds on both the host PC
and the MUD since the game starts. At the end of the experiment, we the 
average SSIM score for each batch of 150 frames and the average throughout
the whole 10-minute experiment as Section~\ref{sec:measure}.

\OpenUVR{} obtains the best SSIM value for a batch at 0.961 and
an average of 0.918 among all batches in our experiments. The average SSIM score
indicates that \OpenUVR{} achieves ``good'' visual quality that surpasses
the 0.9 SSIM threshold~\cite{kahawai}. The SSIM of the worst-case batch in our
experiments is 0.679. The worst-case scenario occurs when the user avatar
enters a new room that is significantly different from the previous room.
We found that the frame rate of the host PC also drops at the same time, a
potential source that affects the efficiency of \OpenUVR{} host system
modules and causes drops of the encoded frames. 

\subsubsection{User study}
\label{sec:user_study_sub}
We also conducted a single-blind A/B user study to identify perceived
differences between an \OpenWVR{} experience and a
native host-PC experience. We randomly selected one hundred college participants across different
institutions. We asked the participants, aged 20--27, to
play two sessions of Unreal Tournament. For session A,
each user played on a direct HDMI
connection to the host PC. For session B,
each user played on an RPi4b using \OpenWVR{}. Both connections were output to 
a single monitor. We used an HDMI switcher to facilitate switching
between the two inputs and set the monitor resolutions to 1920x1080 for both 
sessions. Whether the user started with session A or B was
randomized by flipping a coin. After completing both sessions, we asked the user
which session the user preferred in terms of latency and video quality.

Whereas 51\% of users preferred the \OpenWVR{}-based session on the RPi4b (session B),
49\% preferred the session on the host PC (session A). Most users
stated that they could not identify a difference
between the two sessions, but 24\% of the participants were adamant about either
the host PC or the RPi4b providing a 
substantially better-quality gaming experience.
These results suggest that there is no
perceivable difference between the quality of a
wired system and a well-functioning \OpenWVR{} system.

\subsection{Synchronous \OpenWVR{}}
In its default configuration, \OpenWVR{} initiates an independent thread
that uses a specific GPU-memory location to continually encode data and transmits the data to the 
MUD.
%\CMFSRcom{Does the GPU-memory location encode data and transmit the data to the MUD? If 
%so, then use ``transmit'' in this sentence. If not, the thread first uses the GPU-memory 
%location, and then the thread transmits the data to the MUD, so use ``transmits'' in this sentence.}
This design means that \OpenWVR{}'s encoding and communication processes can pipeline 
with VR-content generation, improving the throughput of system resources.

The default \OpenWVR{} configuration also allows the host
to generate content at high frame rates (90 FPS for Unreal Tournament).
But the Raspberry Pi MUD can only decode and display video at 60~FPS, so generating
VR content at 90~FPS does not improve UX. An application that can render at 90~FPS takes at 
most 1/90th of a second, or 11.1~ms, to execute its game loop. If the application runs
at 60~FPS instead, it has 16.7~ms to run its game loop, giving it 5.6~ms of
slack time to perform \OpenWVR{}'s tasks, which only take around 3.72~ms.

\OpenWVR{} can thus provide a synchronous API that a game can use to trigger 
encoding and data-exchange tasks. The lack of need for
a separate thread also frees up the full CPU core, potentially lowering power 
consumption and/or
the cost of the host PC. In our experiments,
we did not observe any significant change in visual latency or quality due to 
\OpenUVR{}'s default configuration. 

%\input{optimizations}
%\input{user_study}
%\input{discussions}
%\vspace*{-0.1in}
\section{Related Work}
\label{sec:related_works}
As we discussed in Section~\ref{sec:baseline}, the two major bottlenecks of
datapath latency are the video codec and the network software stack. To reduce 
video latency, parallel video codecs~\cite{4285040, 6327343, 6288186, 7892357} 
divide an image into multiple segments to allow parallel encoding/decoding. 
Furion~\cite{Furion} goes further by examining the characteristics of VR content 
to achieve segmentation more efficiently. \OpenUVR{}'s latency reduction techniques, 
which avoid transcoding and virtual-memory overhead, complement parallel video codecs 
and Furion's enhancements of them. In addition, \OpenUVR{} reduces computational 
and energy requirements for VR applications. 

%However, without a full-stack examination,
%Furion and FlashBack both suffer from latencies in each layer of their system elements.

This paper did not explore optimizations for VR-content generation because
other researchers have already succeeded in this area; however, the parallelization 
strategies of VR graphics engines do offer opportunities for improvement, notably with 
respect to picture-frame tiling. The graphics pipeline can be optimized by detecting 
redundant, identical tiles~\cite{RenderingElimination, PIMVR, ShaderMemoization}, by separating stable
objects from frequently updated objects~\cite{10.1145/1450579.1450605,
10.1145/3209661}, or by prioritizing image areas that attract user attention~\cite{8824804, 8816847}. 
Rollback-free value prediction~\cite{ApproximateLoads}, Potluck~\cite{Guo:2018:PCA:3173162.3173185}, 
AxGames~\cite{AxGames}, and PATU~\cite{8327022} apply 
approximate computing techniques to image elements that do not significantly affect
UX. \OpenUVR{} can work with any of these frameworks to reduce the encoded video-stream volume. 

Slow-motion VR applications benefit from strategies that use free bandwidth to aggressively predict, 
precompute, or cache all possible images that a VR user might encounter. As these strategies are 
hardware-independent, many VR systems use them to improve UX. Examples of such systems include
FlashBack~\cite{FlashBackBoos}, CoIC~\cite{CoIC},
Coterie~\cite{10.1145/3373376.3378516}, 
DeltaVR~\cite{10.1145/3302506.3310385}, MUVR~\cite{MUVR}, and the system developed by Mahzari 
et. al~\cite{10.1145/3240508.3240680}. Nonetheless, for fast-moving real-time VR
games, such strategies may not work well since the user's motions continually change. 
Problems are especially acute for VR games with multiple players.

Placing computation-intensive tasks in the cloud provides an alternative VR architecture
for thin user clients~\cite{HaynesVisualCloud,CloudGaming, KamarainenCloudGaming,
Lee2015OutatimeUS,GamingAnywhere}.
VisualCloud~\cite{HaynesVisualCloud} tries to predict
the user's orientation in order to reduce the latency associated with getting data from the cloud,
and VisualCloud reduces bandwidth demand by appending video segments to the media stream. Even this 
combined approach isn't enough to overcome the long
propagation delay between cloud servers and the user device, so state-of-the-art
solutions still suffer from more than a 70~ms delay (and Outatime is no exception).
That being said, some promise may be found among VR applications 
that use edge nodes on 5G networks~\cite{MC}.

As network latency creates serious limitations for
cloud-based solutions, VR systems
using edge-computing architectures are
attracting more attention~\cite{MC,MobileEdge,MSVR,10.1145/3307650.3322264,10.1145/3304109.3306217}.
But these architectures need to distribute workloads between cloud and edge
nodes, which creates task-scheduling issues.
Although communication-constrained mobile-edge computing (MEC)~\cite{MobileEdge} optimizes task
scheduling to minimize high communication costs and maximize
system cache utilization, and although MSVR~\cite{MSVR} hides the long network round-trip
latency by processing delay-sensitive actions on the mobile device, both
architectures require the client or HMD to have a certain degree of
computation capability, which complicates application design. (Note that
Kahawai~\cite{kahawai} presents a novel way to distribute tasks
between the host PC and mobile devices in order to accelerate graphical rendering and
reduce bandwidth.)
Since \OpenWVR{} leaves a relatively small workload to the
client and uses a more powerful
local host PC to generate VR content, \OpenWVR{}'s architecture better
addresses the network delay issue while keeping application design
simple. The demands that \OpenWVR{} places on the MUD are also the lowest among these frameworks.

\OpenWVR{} builds upon previous
research to bypass operating-system overhead and avoid unnecessary memory copies
~\cite{10.5555/1268299.1268320, 10.5555/1268299.1268315, Zerocopy,PerceptionAware,
ShaderMemoization,ApproximateLoads}. Accomplishments from GPUnet can also help reduce 
latencies further~\cite{GPUnet}, but GPUnet's reliance on RDMA hardware
support in wireless NICs limits GPUnet's usefulness
when it is used with existing commodity hardware. Without \OpenWVR{}'s full-stack design,
GPUnet's optimizations can only impact the performance of a single
system component and so fail to eliminate
functionality in each stack layer.

\OpenUVR{}'s network stack resembles user-space
Wi-Fi~\cite{8589152} in that the stack exposes the device layer directly to the
application. \OpenUVR{} adds a kernel module alongside the existing
network interface. This feature allows conventional latency-insensitive applications 
to access the network stack and benefit from optimizations in
each protocol; only \OpenUVR{}DPP traffic goes through the direct network-access 
feature. Note that \OpenUVR{} preserves access to the
network-device buffer as a protected memory address in user-space and
requires the user program to grant such access. Consequently,
\OpenUVR{} does not create security issues in VR systems.  

%With technologies boosting the performance of VR content generation on
%modern GPUs, the proposed \OpenWVR{} architecture becomes feasible as the
%host PC can deliver the desired quality for VR applications.
\ignore{
Generating VR content that \CMFdel{mimics real-life scenarios}\CMF{is realistic}
creates high demand in texture units. Some studies suggest
\CMFdel{the usage of}\CMF{that} approximate
computing \CMF{can accelerate GPU-based generation of VR content on the host PC by using}\CMFdel{to reuse}
\CMFdel{the }similarit\CMFdel{y}\CMF{ies} among picture frames\CMFdel{ to accelerate the GPU
VR content generation on the host PC}~\cite{}.
Approximate deduplication techniques can also be applied to the end-user
program in \OpenWVR{} to further improve \CMFdel{the }latency\CMFdel{ as well as the}\CMF{ and}
bandwidth demand~\cite{GuoPotLuck}.}

Much potential still exists for
architectural and system-level optimizations in VR gaming applications.
Atomic Quake, which demonstrates the usefulness of
transactional memory, serves as one such example~\cite{ZyulkyarovAtomicQuake,QuakeTM}.
Aside from Atomic Quake, the previously mentioned graphical pipeline optimizations, and
\OpenWVR{}, we have found only limited discussion of how to modify
computer architectures to enhance the performance of a complete VR system.

%\vspace*{-0.1in}
\section{Conclusion}
\label{sec:conclude}
Due to severe constraints in latency and
bandwidth, modern UVR systems do not provide
excellent UX.
In addition, existing VR-related research projects only
take single-component approaches
to optimizing system design,
which limits the effectiveness of performance
improvement. Moreover, a cross-layer design is difficult to implement
because many existing VR systems use proprietary
software/hardware components with only partially accessible
designs, thereby discouraging or preventing the research community
from completing a thorough
examination of whole systems.

This paper presents \OpenWVR{}, the first full-system, open-source UVR framework.
\OpenWVR{} works on top of commodity hardware components, making it easy to replicate the \OpenWVR{}
system and study the impact of each architectural
element on the performance metrics to which VR applications
are sensitive. \OpenWVR{} also allows the research community to revisit and evaluate
the interactions among different architectural and system components.
Perhaps most importantly, the UX resulting from \OpenWVR{}'s optimized system architecture
is better than those of existing commercialized products.

Our research shows that (1) a significant amount of latency resides in each
layer of a modern VR system and (2) optimizing system
software stacks and efficiently using architectural components can 
improve end-to-end latency to 14.32~ms, which is imperceptible to the user.
We hope that \OpenWVR{}'s novel open-source design will
enable more research in VR system performance and modification.

%\newpage
%\input{optimizations}
\section*{Acknowledgments}
%\begin{acks}
The authors would like to thank the anonymous reviewers
for their helpful comments. We also owe a debt of gratitude to 
Christopher Fraser for his
excellent copyediting skills. 
This work was partially sponsored by the two National Science Foundation (NSF) awards,
1940048 and 2007124.
This work was also supported by new faculty start-up funds from North Carolina
State University and University of California, Riverside. 
%\end{acks}

  %%%%%%%%%%%%%%%%%%%%%%%%5
%\clearpage
%\pagestyle{empty}
%\input{department}
%\clearpage

%\setstretch{1}
%\bibliography{ws1,ws2,asplos,fpga,full,hot-chips,immd4,microchip,other,BuildingStuff}

%
%\titlespacing*{\section}{0em}{1ex}{1ex}
%\titlespacing{\section}{0em}{1ex}{1ex}
%\begin{small}

%\bibliographystyle{ieeetr}
%\begin{spacing}{0.3}
%\vspace{-0.05in}
\bibliographystyle{ieeetr}
%\begin{spacing}{0.3}
%\bibliographystyle{mbt_dj}
%\scriptsize
%\setstretch{0.82}
%\small
%\fontsize{7.7}{0.0}\selectfont
\bibliography{paper}
%\end{spacing}
%\end{small}

%\clearpage
%\input{bio}
%\clearpage
%\input{budget}
%\clearpage
%\input{facilities}

\end{document}